\documentclass[9pt,twocolumn,twoside]{gsajnl}

\usepackage{epstopdf}
\usepackage{float}
\usepackage{textgreek}
\usepackage[section]{placeins}
\usepackage{gensymb}
\usepackage{siunitx}
\articletype{inv} 

\runningtitle{Literature Review} 
\runningauthor{Jittprasong}

\title{Artificial Intelligence and Medicine: A literature review}

\author[1,$\ast$]{Chottiwatt Jittprasong} 

\affil[1]{Biomedical Robotics Laboratory, Department of Biomedical Engineering, College of Engineering, City University of Hong Kong}

\correspondingauthoraffiliation[$\ast$]{Corresponding author: Biomedical Robotics Laboratory, Department of Biomedical Engineering, College of Engineering, City University of Hong Kong, Kowloon, Hong Kong. \href{chottiwatt.j@my.cityu.edu.hk}{chottiwatt.j@my.cityu.edu.hk}}

\begin{abstract}
In practically every industry today, artificial intelligence is one of the most effective ways for machines to assist humans. Since its inception, a large number of researchers throughout the globe have been pioneering the application of artificial intelligence in medicine. Although artificial intelligence may seem to be a 21st-century concept, Alan Turing pioneered the first foundation concept in the 1940s. Artificial intelligence in medicine has a huge variety of applications that researchers are continually exploring. The tremendous increase in computer and human resources has hastened progress in the 21st century, and it will continue to do so for many years to come. This review of the literature will highlight the emerging field of artificial intelligence in medicine and its current level of development.  
\end{abstract}


\dates{\rec{01 05, 2022} \acc{01 05, 2022}}

\begin{document}

\maketitle
\thispagestyle{firststyle}
\vspace{-13pt}

\section{Introduction}
Artificial Intelligence, which includes Machine Learning and Deep Learning, is the field of study that is concerned with the development of machines that can perform functions that would normally require human intelligence, such as visual perception, speech recognition, decision-making, and language translation between different languages \citep{RN164}. One of the aims of artificial intelligence is to develop machines that can perceive, learn, reason, and behave in the same way humans do. In the last several years, the area of Artificial Intelligence has seen tremendous growth. Financial services, education, and medical are just a few of the industries that have benefited from their utilization. It is also being utilized to develop robots that are capable of doing a wide range of tasks.

Without question, medicine is one of the primary industries that would gain significantly from artificial intelligence integration. The analysis is a critical element of medicine, which is often conducted by physicians, who are also human beings. Unlike humans, artificial intelligence is capable of rapidly processing large volumes of data. While humans are limited to focusing on one or two tasks at a time, artificial intelligence is capable of analyzing massive amounts of data and discovering underlying patterns. Two further features of artificial intelligence are machine learning and adaptation to new environments. Additionally, it is capable of learning from the faults of other artificial intelligence systems and applying that information to future performance improvements. 

The present research trend suggests that artificial intelligence may be used in a variety of medical applications \citep{Buch143}. The use may vary from basic biological data measurement to data analysis and disease diagnosis. 

From this review of the literature, it can be concluded that the use of artificial intelligence in medicine is always evolving toward a more specialized use rather than a more general application. Specifically, numerous publications in the research demonstrated that they modified or developed new machine learning algorithms for their own purposes. As a result, future research into the application of artificial intelligence in medicine is expected to be more focused on the development of novel algorithms for specific medical problems than on the development of a general-purpose AI system. It is also feasible that future development will be more patient-centered than physician-centered, implying that the objective of AI will be to assist the patient rather than to alleviate the physician of duty.

\section{Cardiopulmonary}

\subsection{Computational repurposing of therapeutic small molecules from cancer to pulmonary hypertension \citep{RN153}}

The research developed a differential dependency network (DDN) based on EDDY (Evaluation of Differential DependencY) named "EDDY-CTRP-PH" to connect genes in cancer cells that were associated with pharmacological responses and overlapped with the condition of pulmonary hypertension. The network is being used to reclassify cancer medication that may be used for noncancerous diseases.

EDDY-CTRP-PH was used to assess PH-related gene clusters to identify clusters highly correlated with a cancer cell's responsiveness to medicines and mediators. Cell lines are classified as drug-sensitive or drug-resistant for each cancer medication examined. Within each DDN, clusters of PH-relevant genes that exhibited substantial DDN rewiring across sensitive and resistant cell lines may be candidates for drug repurposing. Additionally, critical network genes called "mediators" are being studied. The rewiring of DDN and the identification of mediators aid in discovering previously undiscovered effects of medications that can mediate the treatment of various disorders.

\begin{figure}[H]
\includegraphics[width=\linewidth]{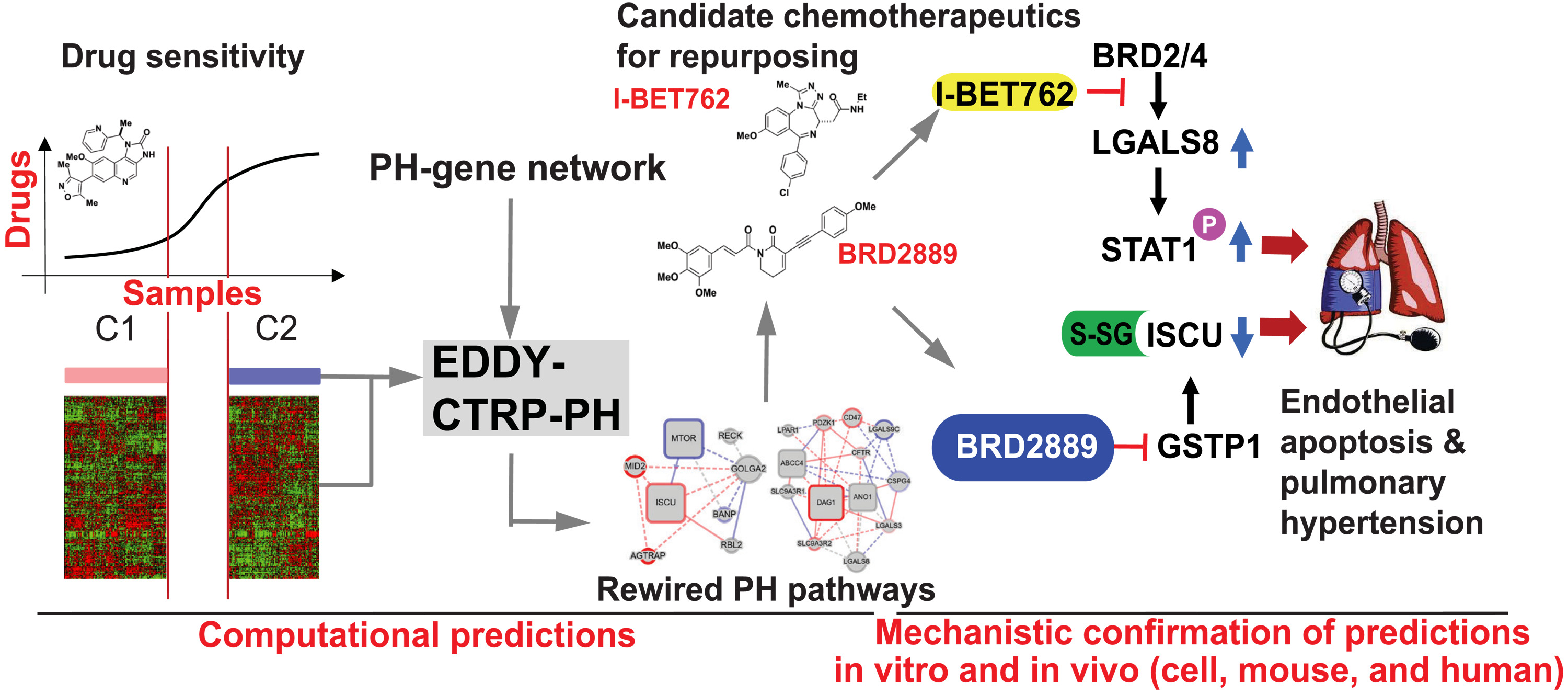}
\caption {EDDY-CTRP-PH: A computational tool for mapping the landscape of cancer drug mechanisms in uncommon noncancerous diseases such as PH. \\
Note: From "Computational repurposing of therapeutic small molecules from cancer to pulmonary hypertension," by \cite{RN153}, \textit{Science Advances}, 7(43). Copyright 2021 by Negi et al. under CC BY license. Reprinted under license terms.}
\label{fig:}
\end{figure}

Because mice were used in the experiment, the availability of littermate controls is limited due to the limited number of mice available for breeding. 

In this study, BET inhibitors were shown to be associated with four rewired PH DDNs that aid in the treatment of PH, and the genes BRD2889, GSTP1, and cluster 43 gene ISCU were discovered to be associated with and capable of improving pulmonary function and arterial hypertension in mice. 

According to the study, a critical drug-gene axis crucial to endothelial dysfunction was identified, as were treatment goals for patients with PH. These findings offer a broad-based, network-dependent framework for redefining cancer medications for use in noncancerous diseases, which may be useful in the future.

\subsection{Dynamic detection and reversal of myocardial ischemia using an artificially intelligent bioelectronic medicine \citep{RN136}}

This research is a proof-of-concept of an Artificial Neural Network (ANN) supplementing the Myocardial Sensory Networks in reliably detecting Myocardial Ischemia through the decoding of cardiovascular pathophysiological features along with the reversal of the Myocardial Ischemia by ANN-controlled Vagus Nerve Stimulation (VNS).

The ANN architecture is composed of a 4-layer (sequence input, dense layer, LSTM layer, and class layer.) The Long short-term memory (LSTM) layer is incorporated to enable the ANN an ability to detect long-term dependencies of cardiovascular data, increasing the likelihood of recognizing broad features of Myocardial Ischemia. The architecture is trained via supervised learning to equip the ANN with the ability to differentiate states of induced-Myocardial Infarction (4 states: rest, dobutamine infusion: D, norepinephrine infusion: NE, and D + NE). Recognition of cardiovascular stress by the ANN is determined through observing broad sets of features of cardiovascular physiological parameters (e.g., ECG and BP). The output of the ANN, determined by a decoder output score of > 0.5, will trigger delivery of Closed-loop intact Vagus Nerve Stimulation, thus, reversing all major physiological features of induced Myocardial Infarction with more probability of success and low side effect incidence rate.

\begin{figure}[H]
\includegraphics[width=\linewidth]{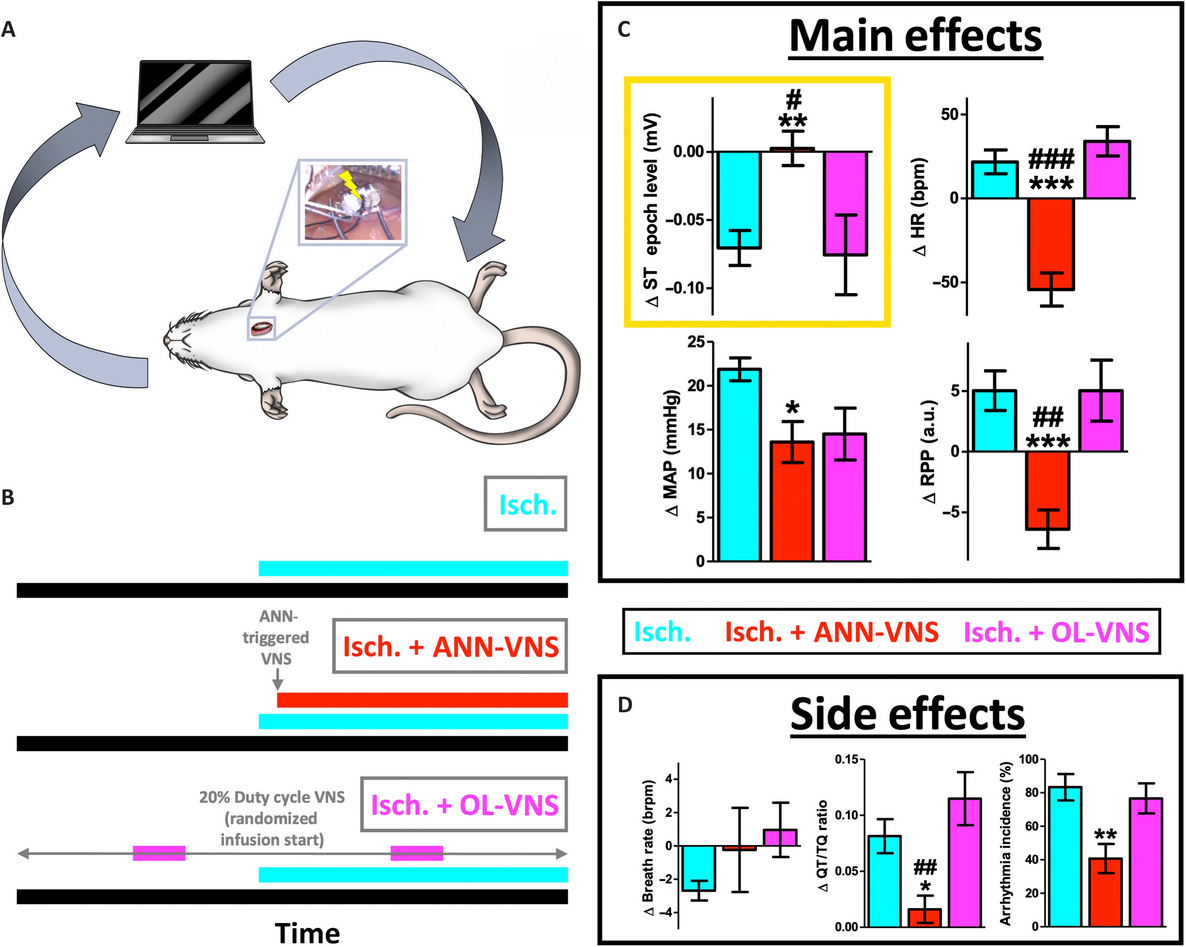}
\caption {\textbf{(A)} The ANN was employed in vivo to identify and control spontaneous myocardial ischemia (ANN-VNS; inset: left cervical vagus nerve and VNS cuff during implant). \textbf{(B)} The article examined D + NE ischemia (cyan, Isch.), D + NE ischemia and closed-loop ANN-VNS (red, Isch. + ANN-VNS), or D + NE ischemia and open-loop VNS (magenta, Isch. + OL-VNS). \textbf{(C)} Closed-loop ANN-VNS (red, Isch. + ANN-VNS) corrected various physiological aspects of myocardial ischemia, including ST epoch level, heart rate, MAP (a correlate of afterload), and RPP (an index of myocardial oxygen consumption). Ischemia with open-loop VNS (magenta, Isch. + OL-VNS) failed to correct physiological aspects of myocardial ischemia and was virtually no different from Ischemia alone (cyan, Isch.) \textbf{(D)} The rate of breathing did not differ across groups. Only closed-loop ANN-VNS significantly decreased the probability of reentry (QT/TQ ratio) and arrhythmia incidence. These results suggest that artificial neural networks (ANNs) may supplement myocardial sensory networks and aid in the reversal of spontaneous cardiac ischemia through bioelectronic treatment. \\
Note: From "Dynamic detection and reversal of myocardial ischemia using an artificially intelligent bioelectronic medicine," by \cite{RN136}, \textit{Science Advances}, 8(1). Copyright 2022 by Ganzer et al. under CC BY license. Reprinted under license terms.}
\label{fig:}
\end{figure}

However, the ANN still suffers from common problems of neural models, such as outdated models. Therefore, over time and without continuous learning, it will eventually fail to recognize new physiological changes resulting from new kinds of stresses. Additionally, the research is still only a pre-clinical study.

In the implementation, 13 pathophysiological features were chosen for cardiovascular state decoding via ANN. ECG and BP were determined essential parameters to cardiovascular state decoding accuracy. ANN-controlled VNS was found to reverse most pathophysiological features compared to others significantly. Implementation of LSTM autoencoders enables ANN to detect new stress states with high sensitivity.

\section{COVID-19 Pandemic}

\subsection{Artificial Intelligence and COVID-19: deep learning approaches for diagnosis and treatment \citep{RN141}}
The paper assesses several conceptual structures of the AI-based techniques to assist in COVID-19 diagnostic and treatment protocols. Various utilization of different types of Artificial Neural Networks (ANN) is designed to help facilitate the conventional methods of diagnosing the COVID-19.

The ANN-based selector to choose the most appropriate conventional medical imaging techniques (such as MRI, CT, X-ray, PET) that has a probability of displaying the manifestation of COVID-19 infection. The imaging results are then inputted into ANN-based image processing to optimize and then classify the diagnosis by ANN-based estimator. Extreme Learning Machine (ELM) ANN was designed and studied to predict suitable drugs and estimate the hospital stay and symptomatic period. Long/Short-Term Memory (LSTM) can be designed to classify the best treatment option given biomarkers. A recurrent Neural Network (RNN) with real-time learning can be utilized to predict the outbreak. Generative Adversarial Network (GAN) can be clinically applied along with Transmission Electron Microscopy to detect the infection of COVID-19.

\begin{figure}[H]
\includegraphics[width=\linewidth]{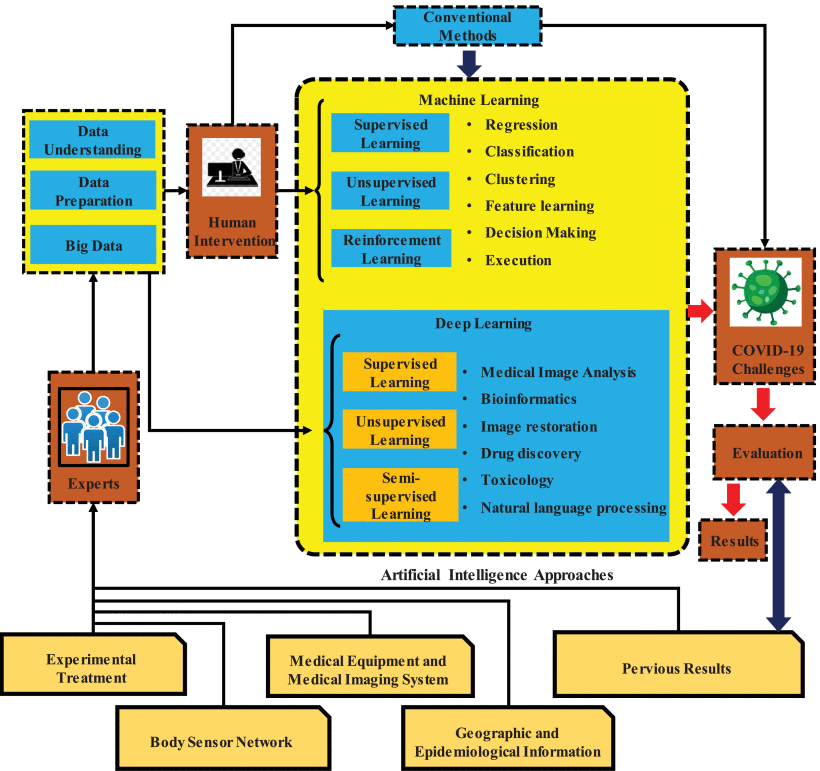}
\caption {The methodology diagram of using AI-based approaches to overcome COVID-19-related issues.  \\
Note: From "Artificial Intelligence and COVID-19: Deep Learning Approaches for Diagnosis and Treatment," by \cite{RN141}, \textit{IEEE Access}, 8, 109581-109595. Copyright 2020 by Jamshidi et al. under CC BY license. Reprinted under license terms.}
\label{fig:}
\end{figure}

In this paper, it was found that many ANN-based techniques have been revealed a potential to assist in the diagnosis and treatment of COVID-19 and a supervised learning approach with AI and physicians complimenting the diagnosis and treatment options of COVID-19 is one of the optimal way. However, proposed techniques have not been applied in the actual settings to evaluate their effectiveness and limitations. Experiments with real dataset are necessary to optimize the use.

\subsection{COVID-19 Artificial Intelligence diagnosis using only cough recordings \citep{RN143}}
The research highlights the usage of AI-based COVID-19 diagnosis using cough recording analysis, which may compensate for the shortcomings of the current gold standard for COVID-19 detection, Realtime RT-PCR. Cough recordings have been shown to be effective, quick, simple to establish, non-invasive, and inexpensive. This qualifies individuals for mobile or remote diagnostics, which may result in fast referral for diagnosis and treatment.

The study team gathered varying lengths of cough audio recordings (on average, three coughs per person) and ten multiple-choice questions about illness diagnosis. The proposed architecture takes a recording containing one or more coughs, conducts two pre-processing steps on it, and then feeds it into a CNN-based model to provide a pre-screening diagnostic as well as a biomarker saliency map. The AI biomarker architecture is then used to analyze four biomarkers (muscular deterioration, vocal cords, sentiment, and respiratory tract). A Poisson mask is applied to the cough recording to assess muscle degradation. ResNet50 has been taught to distinguish aspects of a person's vocal cord using COVID-19. It is also used to recognize a patient's emotions and conditions of the respiratory tract, which may indicate a probable infection. Transfer learning also showed a possible improvement in detection in other languages. Global Average Pooling 2D layer and 1024 neural network layer with ReLU are then used to designate.

\begin{figure}[H]
\includegraphics[width=\linewidth]{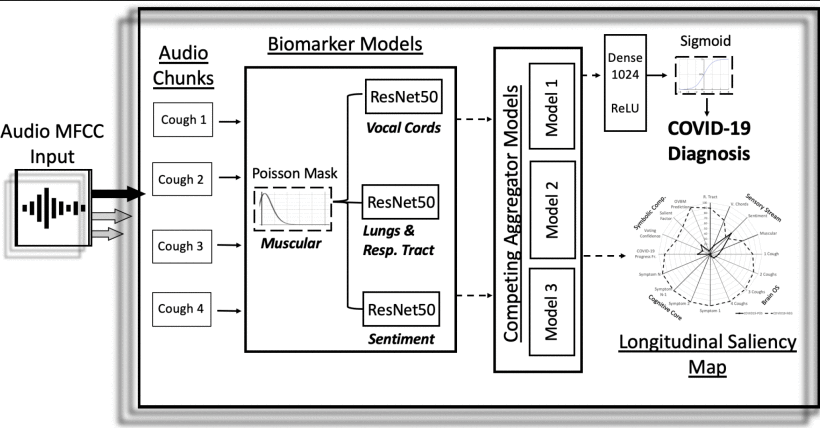}
\caption {Overview architecture of the COVID-19 discriminator with cough recordings as input, and COVID-19 diagnosis and longitudinal saliency map as output.\\
Note: From "COVID-19 Artificial Intelligence Diagnosis Using Only Cough Recordings," by \cite{RN143}, \textit{IEEE Open Journal of Engineering in Medicine and Biology}, 1, 275-281. Copyright 2020 by Laguarta et al. under CC BY license. Reprinted under license terms.}
\label{fig:}
\end{figure}

The limitation was that the proposed protocols did not evaluate cultural, language, and age differences which may affect input modalities and thus resulted in deviated diagnostic results.

It was found that the cough biomarker model significantly improved the discrimination accuracy in diagnosing COVID-19. This showed that the cough pre-screening tool might be used to rule out a group of patients under investigation (PUI), thus, avoiding excessive wide testing. Additionally, it was found that multiple clinical manifestations designated as discriminatory biomarker values of COVID-19 and Alzheimer's disease are the same. 

\section{Imaging Systems}

\subsection{Accurate auto-labeling of chest X-ray images based on quantitative similarity to an explainable AI model \citep{RN163}}

The article established an explainable artificial intelligence model, called xAI, that can be used to reliably classify chest X-ray images based on their quantitative similarity to the training set. The model is completely adjustable, and tuning the original model, enables performance retraining, preservation, or enhancement of the model.

xAI can compute the probability-of-similarity metric pSim, which is a value used to estimate the likelihood of similarity between the input and the model's reference atlas, using a model-derived atlas-based approach. The resulting confidence and patch similarity calculations are used to calculate the pSim value based on the harmonic mean of the two before deciding whether to label the input according to the confidence threshold automatically or to notify when the pSIm value goes below the threshold. This approach also allows for a fine-tuning of the original model to retrain the model to detect other features to be applied to a different external dataset.

\begin{figure}[H]
\includegraphics[width=\linewidth]{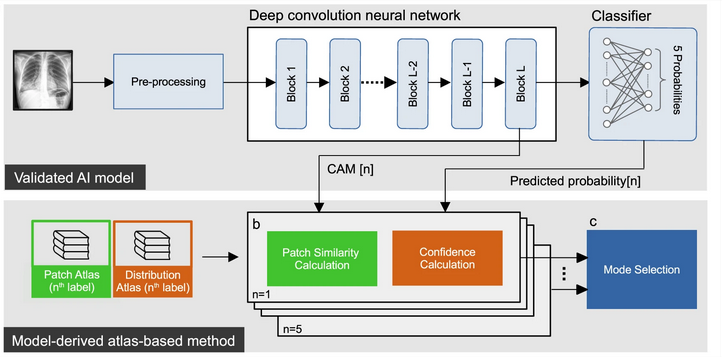}
\caption {Overview of the system. Standardized, automated labeling technique based on similarities to a previously validated five-label chest X-ray (CXR) detection explainable AI (xAI) model, employing an atlas-based methodology developed from the xAI model. An atlas-based explainable AI system derives a pSim value for automatic labeling based on the harmonic mean of patch similarity and confidence. The generated pSim metric may be used in a “mode selection” algorithm to either label or notify the user when the pSim value goes below a threshold.\\
Note: From "Accurate auto-labeling of chest X-ray images based on quantitative similarity to an explainable AI model," by \cite{RN163}, \textit{Nature Communications}, 13, 1867. Copyright 2022 by Kim et al. under CC BY license. Reprinted under license terms.}
\label{fig:}
\end{figure}

The initial training set's quality determines the labeling accuracy and performance with this approach. Therefore, the quality of the original training set may influence the final labeling.

It was concluded that auto-labeled CXR images indicating cardiomegaly and pleural effusion agree with expert radiologists in over 80\% of cases but contradict expert radiologists' labeling of pulmonary edema and pneumonia, which implies that they are non-specific and that different radiologists use different criteria to classify the conditions.

\subsection{A fully automatic AI system for tooth and alveolar bone segmentation from cone-beam CT images \citep{RN162}}
The article developed a completely automated segmentation of cone-beam computed tomography (CBCT) images using a deep-learning-based artificial intelligence (AI) system to differentiate individual teeth and alveolar bones. Automatic area of interest (ROI) identification is used to allow completely automated segmentation. Current techniques, such as ToothNet and CGDNet, need the operator to manually define the ROI, thus, not fully automated.

The input image of 3D CBCT images is normalized to a resolution of 0.4 x 0.4 x 0.4 mm\textsuperscript{3}. After that, the images are sent into the AI system. The ROI for tooth recognition is extracted using a V-Net. Following that, the hierarchical morphology-guided network automatically segments the individual teeth into two outputs: centroid and skeleton. These outputs are then used to define each tooth precisely using the multi-task learning network. The architecture for segmenting alveolar bones using an enhanced neural network is utilized to discriminate between the midface and mandible bones in the input images. The Harr transform is used to enhance the boundaries in the CBCT image before it is sent to a cascaded V-Net to generate a bone mask. The bone mask is then combined with the teeth segmentation network.

\begin{figure}[H]
\includegraphics[width=\linewidth]{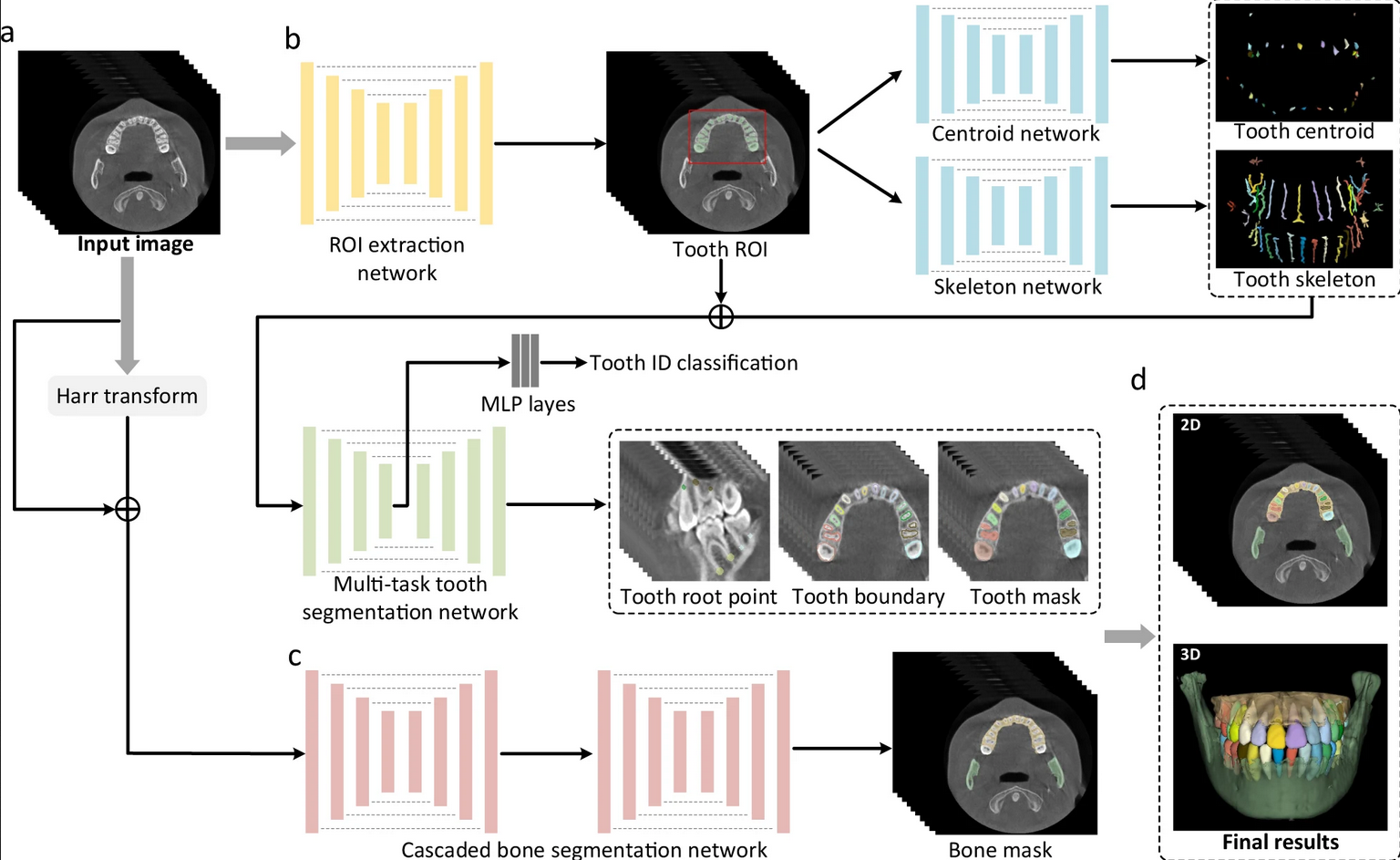}
\caption {A diagrammatic representation of our proposed artificial intelligence system for segmenting individual teeth and alveolar bones from CBCT pictures. \textbf{(A)} The system is fed data from a 3D CBCT scan. \textbf{(B)} Individual teeth are segmented using the morphology-guided network. \textbf{(C)} Alveolar bones are extracted using the cascaded network. . \textbf{(D)} The model generates masks for individual teeth and alveolar bones.\\
Note: From "A fully automatic AI system for tooth and alveolar bone segmentation from cone-beam CT images," by \cite{RN162}, \textit{Nature Communications}, 13, 2096. Copyright 2022 by Cui et al. under CC BY license. Reprinted under license terms.}
\label{fig:}
\end{figure}

In this paper, the model implementation imposed a memory constraint on the GPU, resulting in a need to reduce the resolution of the input images.

It was concluded that the segmentation performance in recognizing abnormalities in alveolar bone and tooth is more than 90\% and the qualitative segmentation by the AI is comparable to the segmentation done by expert radiologists. Moreover, the AI performed better than expert radiologists in delineating teeth, which is relatively small objects.

\subsection{Deep-learning-based identification, tracking, pose estimation and behaviour classification of interacting primates and mice in complex environments \citep{RN161}}

The article established an approach for detecting particular animal behaviors in complicated environments for use in systems neuroscience analysis. They developed a pipeline called SIPEC with numerous series of them for behavioral detection (SIPEC:BehaveNet), segmentation (SIPEC:SegNet), identification (SIPEC:IdNet), and pose estimation based on the deep neural network (DNN) (SIPEC:PoseNet). These four DNNs can be employed directly on target videos.

First, the SegNet, with optimized convolutional neural network (CNN) backbone architecture including Mask R-CNN architecture, ResNet101, and feature pyramid network (FPN), is used to mask and define the bounding box of each target animal. This result is then used to identify the animal species separately using the IdNet, which is based on the DenseNet architecture. After obtaining information from SegNet and IdNet, BehaveNet, which is based on a raw-pixel action recognition network and feature recognition network (FRN) based on Xception, analyzes the target animal's real and usual behaviors to identify any particular behaviors shown by the animal. Additionally, PosNet, based on the encoder-decoder design and EfficientNet, is utilized to determine the animal's markers that define its posture.

Even though SIPEC components are intended to operate in complex environments, they are still susceptible to inaccuracy due to environmental changes such as lightning.

It was concluded that under demanding conditions, the ground truth and prediction of SIPEC's SegNet, PoseNet, and IdNet remain consistent, and SIPEC:BehaveNet outperformed the widely used behavioral detection algorithm DeepLabCut. Additionally, with a single camera, SIPEC:BehaveNet can recognize social interactions among primates inferred to three-dimensional locations.

\section{Oncology}
\subsection{A perception-based nanosensor platform to detect cancer biomarkers \citep{RN147}}

This study aimed to overcome the constraints of one-to-one recognition, which is frequently used in perception-based sensing systems. The study team built a DNA-SWCNT-based (SWCNT: Single-wall carbon nanotube) photoluminescent sensor array that employed machine learning algorithms in conjunction with optical response training to identify gynecologic cancer biomarkers in biofluids.

The sensory array was constructed using DNA-SWCNT complexes composed of eleven DNA sequences wrapped around twelve semiconducting SWCNT species, yielding 132 distinct DNA-SWCNT complex combinations. The optical responses of the DNA-SWCNT complexes suggestive of gynecologic cancer are primarily directed towards the protein biomarkers HE4, CA-125, and YKL-40. The DNA-SWCNT complex incubated with human biofluids then undergoes high-throughput NIR spectroscopy to record their optical responses, where the wavelength and intensity of each sensor band are determined. The data are then processed into a feature vector (FV) and verified by machine learning algorithms (preferentially Random Forest and Artificial Neural network) trained with an FV training dataset to classify each protein, its combinations, and interpretations.

\begin{figure}[H]
\includegraphics[width=\linewidth]{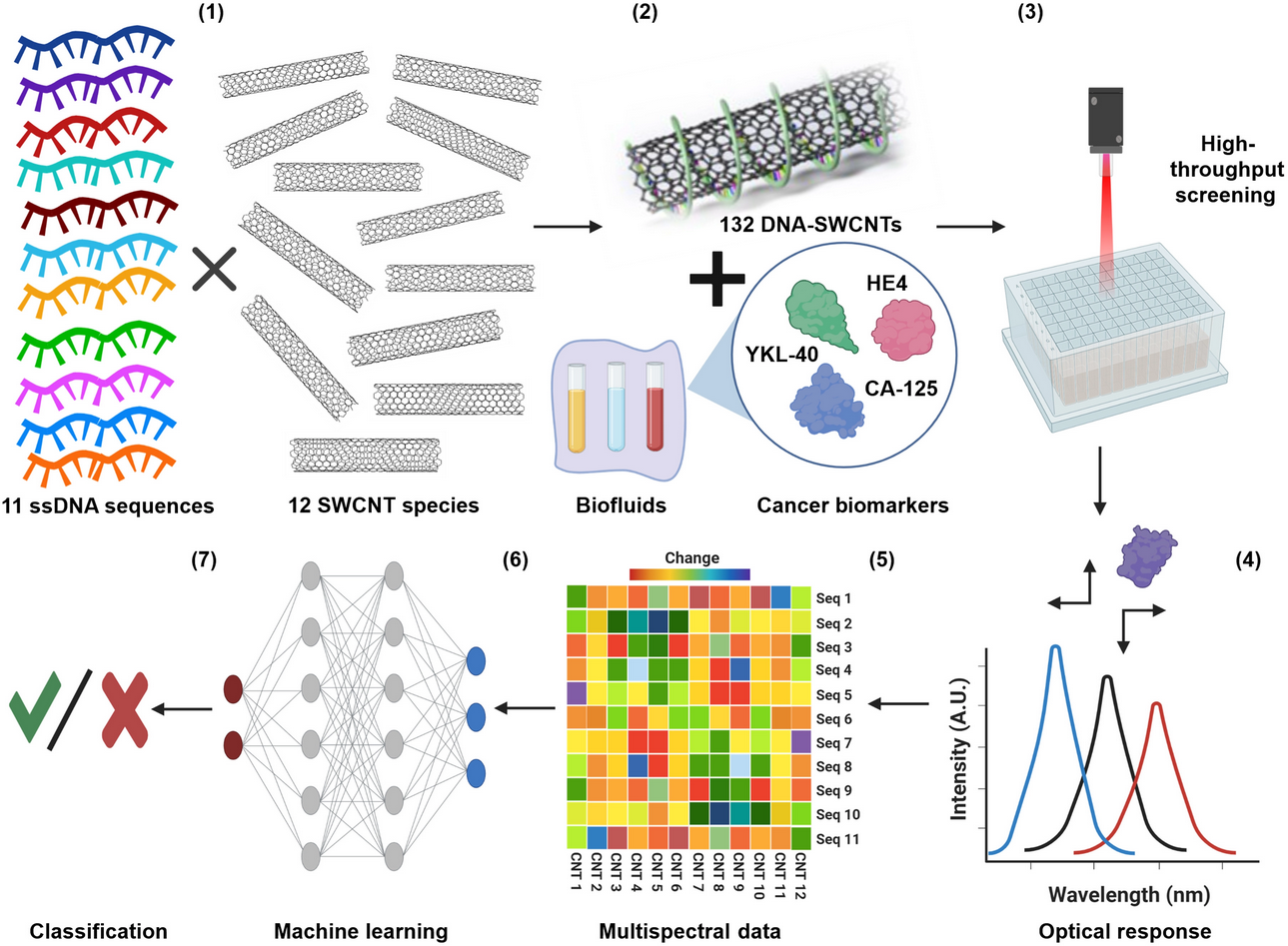}
\caption {Platform for protein biomarker nanosensors.(1) Eleven single-stranded DNA oligonucleotides wrap SWCNT chiralities. (2) The sensor array is incubated in the sample. (3) High-throughput NIR spectroscopy probes the sensors' optical response. Fourier transforms are used to match the spectroscopic data to each sensor emission band. (5) The sensor responses are turned into an FV training set. A.U.,arbitrary units. (6) ML algorithms are developed and tested for each target protein. Seq, sequence; CNT, Carbon nanotubes. (7) Predictions are assessed. \\
Note: From "A perception-based nanosensor platform to detect cancer biomarkers," by \cite{RN147}, \textit{Science Advances}, 7(47). Copyright 2021 by Yaari et al. under CC BY-NC license. Reprinted under license terms.}
\label{fig:}
\end{figure}

The limitation was that interferents might block DNA-SWCNTS response to an analyte in a complicated environment. Trying to separate DNA-SWCNT responses from protein biomarkers and interferents failed.

This paper found that Receiver Operating Characteristic curves revealed significantly high detection accuracy in SVM and RF. A detection threshold less than 100 pM was able to accurately classify biofluid samples.

\subsection{Artificial intelligence for diagnosis and Gleason grading of prostate cancer: the PANDA challenge \citep{RN157}}

The PANDA consortium, which includes thousands of developers, is striving to develop artificial intelligence algorithms for the diagnosis of prostate cancer based on histopathology. The algorithms determine Gleason grading based on digitized prostate biopsies. Across geographic cohorts, the performance achieved pathologist-level performance.

The PANDA consortium, which includes thousands of developers, is striving to develop artificial intelligence algorithms for the diagnosis of prostate cancer based on histopathology. The algorithms determine Gleason grading based on digitized prostate biopsies. Across geographic cohorts, the performance achieved pathologist-level performance. The Kaggle computation environment of the algorithms is validated and run on internal validation sets to validate the performance. After internal sets were run, external validation sets of the US and EU were run with the algorithms to evaluate the performance. After running the internal validation sets, the algorithms were used to assess the performance of external validation sets from the US and EU. Cohen's K values are used for statistical analysis.

\begin{figure}[H]
\includegraphics[width=\linewidth]{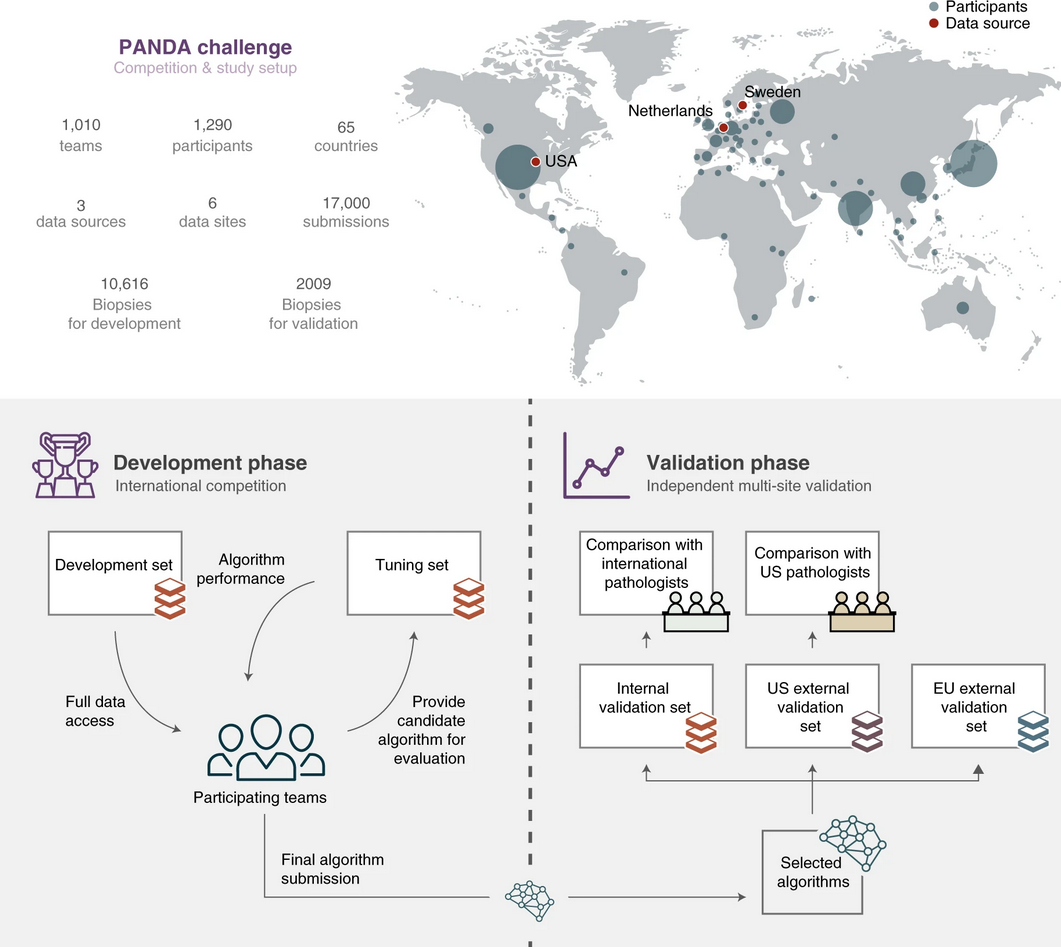}
\caption {Participants came from 65 nations (top: size of the circle for each country illustrates the number of participants). The research has two phases. Phase 1: Teams competed to construct the best-performing Gleason grading algorithm, with full access to the development set for algorithm training and restricted access to the tuning set for algorithm performance estimation. In the validation phase (bottom right), algorithms were tested separately on internal and external datasets against consensus uropathologist grading and compared to groups of foreign and US general pathologists on subsets of data.\\
Note: From "Artificial intelligence for diagnosis and Gleason grading of prostate cancer: the PANDA challenge," by \cite{RN157}, \textit{Nature Medicine}, 28, 154–163. Copyright 2022 by Bulten et al. under CC BY license. Reprinted under license terms.}
\label{fig:}
\end{figure}

The limitation was that the change in data distribution between training and external validation data resulted in an overdiagnosis of a single instance.

In this study, the algorithms performed significantly better than the pathologists with higher sensitivity and specificity, and significant improvement of the algorithms was observed just a week after commencement.

\subsection{Brain tumor detection using statistical and machine learning method \citep{RN139}}
This study focus on finding a new approach to overcome limitations in current medical image analysis methodology and improving its efficiency. The study targets the detection of brain tumors aiming to enhance its efficacy in early detection by implementing enhancement, segmentation, and features fusion process through machine learning. The new approach in the segmentation method and feature vector experiment showed a better result than the current techniques.

The approach starts by using the Weiner filter to minimize artifacts and noises before image processing and enhance the image's detail, a so-called lesion enhancement. Segmentation of the image into slices by Potential field (PF) clustering was then performed to refine for a subset of tumor pixels, including a process of morphological erosion and dilation, fine-tuning the final image prior to feature extraction. The fused feature vector is then obtained by concatenation of Local binary pattern (LBP) and Gabor wavelet transform (GWT) features extracted from the image. Score-dependent feature selection based on Boltzmann entropy is then performed to filter 50 out of 89 features for classification. Various types of classifiers and then used to classify the vector into tumor/non-tumor, such as Support vector machine (SVM) and k-nearest neighbor (KNN), to compare the performance.

\begin{figure}[H]
\includegraphics[width=\linewidth]{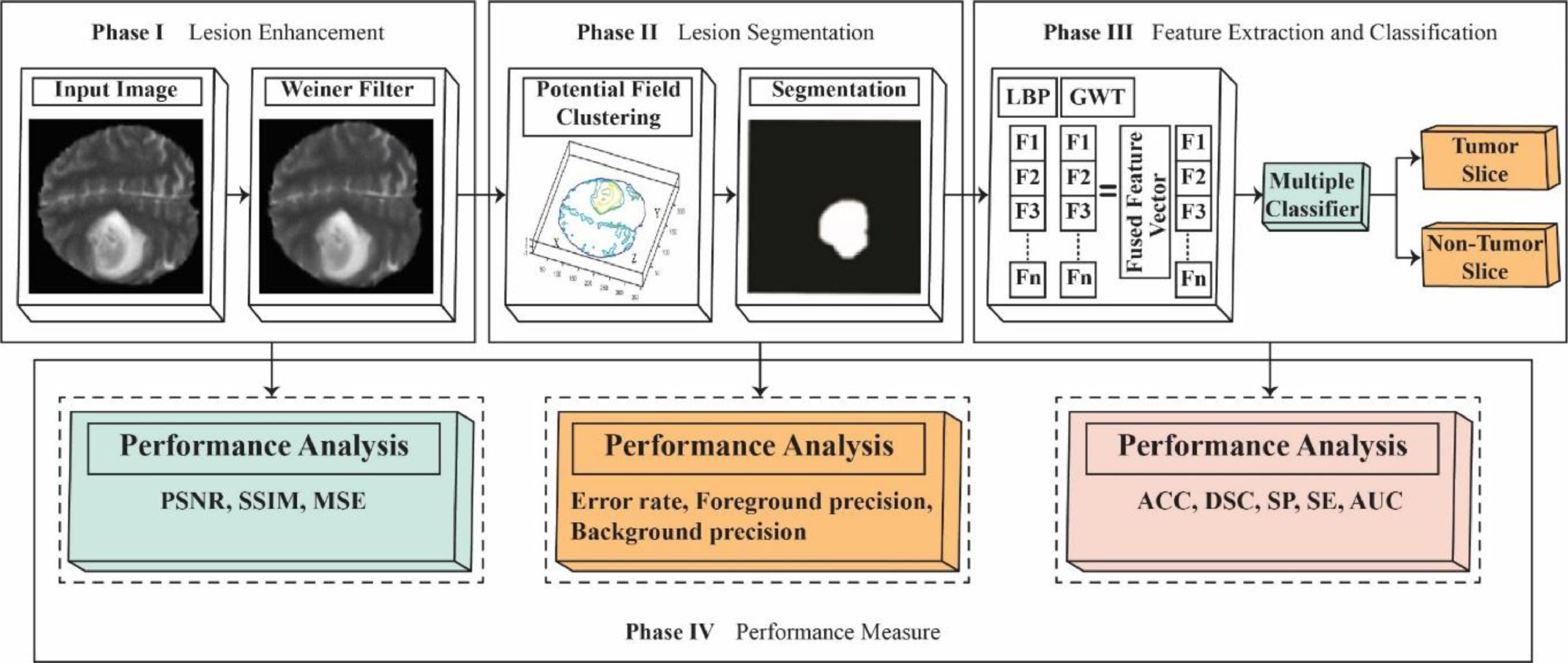}
\caption {Proposed method block diagram \\
Note: From "Brain tumor detection using statistical and machine learning method," by \cite{RN139}, \textit{Computer Methods and Programs in Biomedicine}, 177, 69-79. Copyright 2019 by Elsevier. Reprinted with permission.}
\label{fig:}
\end{figure}

The limitation was that the technique was applied on only one range of datasets (BraTS, a multimodal brain tumor segmentation dataset) and a local dataset; further use with different datasets has not been analyzed.

\subsection{Detection of ovarian cancer via the spectral fingerprinting of quantum-defect-modified carbon nanotubes in serum by machine learning \citep{RN154}}
The work established a technique for detecting ovarian cancer in serum samples by extracting a spectral fingerprint from the near-infrared fluorescence emissions of quantum-defect carbon nanotubes. The algorithm’s performance is slightly better than the current screening test of antigen 125 and transvaginal ultrasonography.

Support vector machine trained to detect abnormal levels of disease biomarkers tuned with Bayesian optimization was used to perform a binary discrimination of high-grade serous ovarian cancer (HGSOC) or other diseases/health based spectroscopic variables. The spectroscopic variables are acquired using a nanosensor array composed of organic colour centres-functionalized single-stranded DNA (ssDNA), encapsulated signal-walled carbon nanotubes (SWCNTs), or an OCC-DNA nanosensor that responds to HGSOC-specific biomarkers (e.g. CA125, HE4, and YKL40) in serum samples. Support vector regression (SVR) model was then used to quantitatively predict biomarker levels using the sensor array.

However, SVR models indicated that the obtained predictive value did not account for all markers, suggesting the presence of undiscovered biomarkers.

It was discovered that the algorithm performed 87\% sensitivity and 98\% specificity in discriminating HGSOC.

\subsection{Lung and pancreatic tumor characterization in the deep learning era: novel supervised and unsupervised learning approaches \citep{RN140}}

The research study the enhancement of computer-aided diagnosis (CAD) tools with two types of machine learning algorithms (supervised and unsupervised learning algorithm) utilizing a 3D convolutional neural network (3D CNN) to perform Intraductal Papillary Mucinous Neoplasm (IPMN) and lung tumor characterization. The unsupervised learning approach is studied based on previous works to overcome the limitation of the supervised learning approach where resources needed to label large datasets are extensive.

In the supervised learning approach, features obtained from n images of lung nodules are given attribute/malignancy scores based on feature representation. A pre-trained 3D CNN fine-tuned with a CT scan dataset (acquired dense feature representation) is then used to obtain a malignancy score from the image along with the Multi-Task Learning (MTL) approach utilized to evaluate the level of dependencies among visual attributes distinctively. In an unsupervised learning approach, features are extracted from input images (lung nodules/IPMN), k-means clustered and given an initial label. Label proportions are then computed into each cluster. A proportion-Support Vector Machine ($\propto$SVM) is then used to study the discriminative model, characterize, and assign malignancy/normal label to each cluster.

The limitation to this approach was that the pre-labeled data used to train the neural network (supervised learning approach) are labeled by humans; unless backed up by biopsy, large variability may exist between different radiologists. 

In this study, it was found that a supervised learning approach with 3D CNN and MTL implemented succeeded other means of characterization in terms of accuracy, and promising results of an unsupervised approach with $\propto$ SVM were obtained compared to other approaches. 

\subsection{Machine learning analysis of DNA methylation profiles distinguishes primary lung squamous cell carcinomas from head and neck metastases \citep{RN138}}

The study developed a machine learning trained to profile a DNA methylation differentiating head and neck squamous cell carcinoma (HNSC) from primary squamous cell carcinomas (LUSCs), which are highly difficult to distinguish, given the current diagnostics techniques and its similar histomorphology and immunohistochemical profiles.

Identification of epigenetic signatures of primary HNSC, LUSC, and control samples was carried out to be used as a reference cohort. Two thousand CpG sites were selected and used as a variable for a classifier based on artificial networks. The classifier is trained with the CpG sites database to determine the probability of the organ of origin of the tumor based on the DNA methylation profile of cancer. Fivefold cross-validation based on the reference cohort was carried out to tune the model. The final model is used in the independent validation cohort of patients to differentiate primary HNSC, LUSC, and control based on DNA methylation profile.

\begin{figure}[H]
\includegraphics[width=\linewidth]{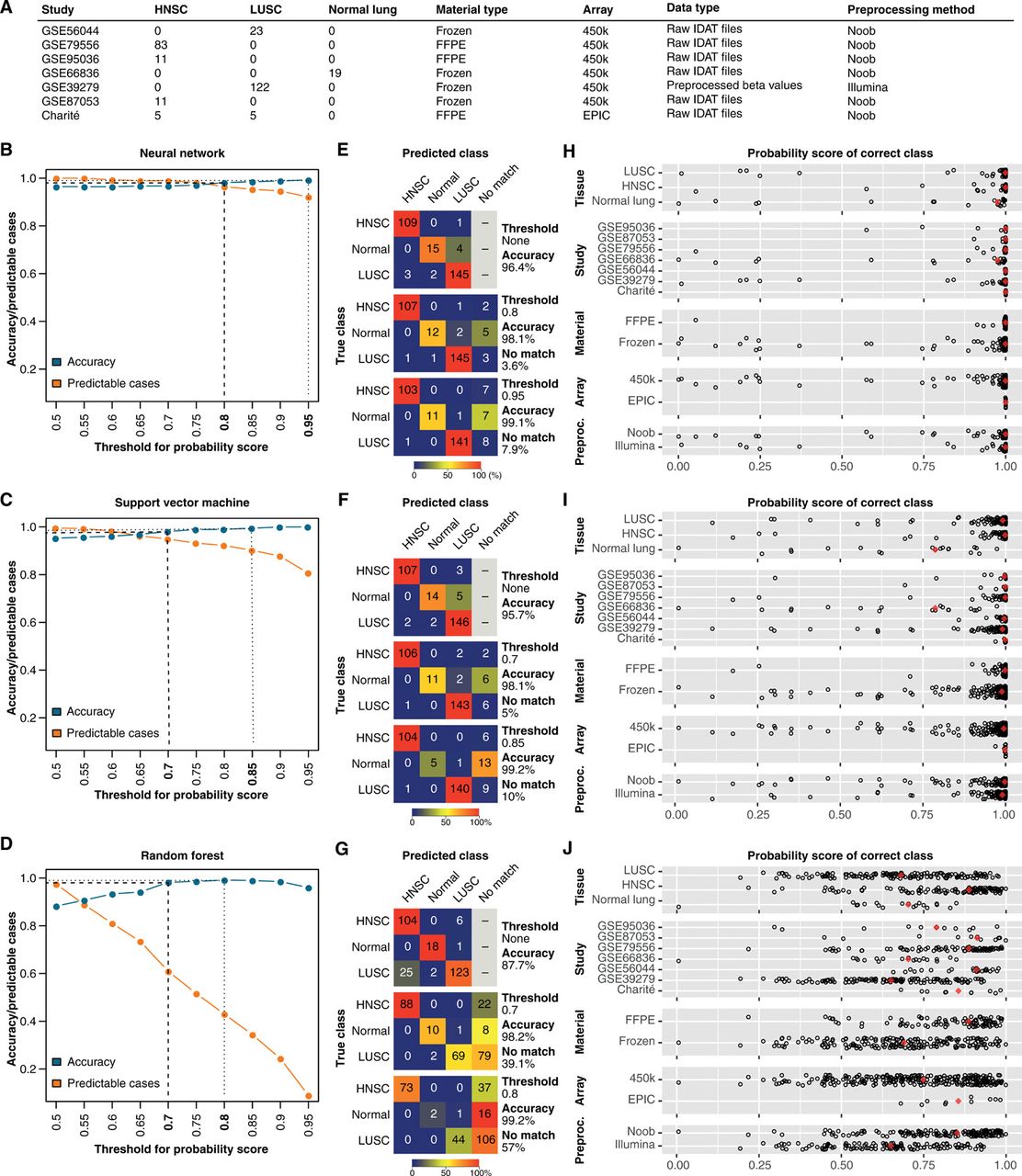}
\caption {Results of machine learning algorithms on a validation cohort. \textbf{(A)} Validation cohort samples (n = 279). \textbf{(B} to \textbf{D)} Probabilistic neural network (B), support vector machine (C), and random forest (D) prediction score threshold analysis. \textbf{(E} to \textbf{G)} Confusion matrices for HNSC (n = 110), LUSC (n = 150), and normal lung tissue (n = 19) samples using neural networks (E), support vector machine (F), and random forest (G). \textbf{(H} to \textbf{J)} Variations in probability scores (range, 0 to 1) for the correct class for neural networks (H), Support vector machine (I), and random forests (J) using various tissues and pre-processing techniques.\\
Note: From "Machine learning analysis of DNA methylation profiles distinguishes primary lung squamous cell carcinomas from head and neck metastases," by \cite{RN138}, \textit{Science Translational Medicine}, 11(509). Copyright 2019 by Jurmeister et al. and American Association for the Advancement of Science. Reprinted with permission.}
\label{fig:}
\end{figure}

The limitation of this study was the research did not investigate the usage and accuracy of the model trained with a particular anatomic site to differentiate carcinomas in the different anatomic sites, which the model was not trained prior. In this case, the model was trained to differentiate only HNSC and LUSC and validated for those two only. Additionally, the current methodology requires precise analysis for accurate prediction.

In this article, it was found that the analysis based on DNA methylation succeeded the mRNA expression and proteomics analysis in prediction accuracy, and DNA methylation analysis via ML yielded superior results compared to SVM and random forests. Additionally, the network may be trained with unlabeled data, which may increase the prediction accuracies and aid in future studies.

\subsection{Metabolic detection and systems analyses of pancreatic ductal adenocarcinoma through machine learning, lipidomics, and multi-omics \citep{RN137}}

This research studies the detection of Pancreatic ductal adenocarcinoma (PDAC) through a combination of Machine Learning, Support Vector Machine (SVN), metabolomics, and lipid metabolites. Differentiation involves the use of an ML-based greedy model where various features of PDAC were chosen, validated, and further filtered by MS-based selection and LC-MS-based assay to distinguish PDAC from Control, ensuring high reliability and accuracy of the ML-aided metabolic PDAC detection approach.

An exploratory study was conducted to collect samples from PDAC patients and normal control (NC) for untargeted lipid metabolomics profiling with data cross-validation. SVM was then applied to classify PDAC from NC with all detected lipids data. The ML-based and greedy-based model was then used to extract “valuable” features from the lipidomics data and rule out “noise,” which cannot be used for PDAC detection. The greedy algorithm will weight the extracted features to rank their importance in which the weighted features will be filtered and refined through MS-based selection again. A total of 17 lipids are then chosen for targeted lipidomics assay (LC-MS-based) to detect PDAC accurately.

\begin{figure}[H]
\includegraphics[width=\linewidth]{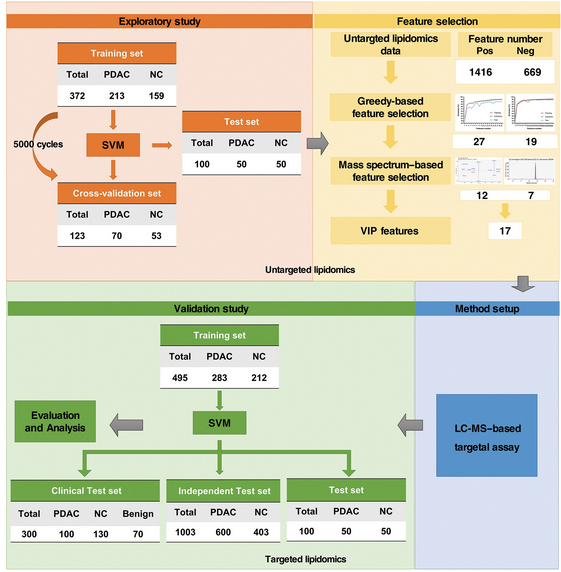}
\caption {Descriptive research design and schematic representation of ML-aided metabolic PDAC detection. An SVM model is set up for an exploratory research (n = 595). The greedy algorithm and LC-MS selected the features. The targeted lipid MRM-mode quantification assay and validation study [n = 1898; 495 for training set, 100 for internal test set, 1003 for independent test set, and 300 for clinical test set] are described. \\
Note: From "Metabolic detection and systems analyses of pancreatic ductal adenocarcinoma through machine learning, lipidomics, and multi-omics," by \cite{RN137}, \textit{Science Advances}, 7(52). Copyright 2021 by Wang et al. under CC BY-NC license. Reprinted under license terms.}
\label{fig:}
\end{figure}

The limitation was that the model used in this study was not trained separately with distinct features of early PDAC and late stages PDAC but only the standard features of both; thus, lacking the ability to determine prognosis and stages.

This article found that the ML-aided metabolic PDAC detection in conjunction with the greedy algorithm and LC-MS feature selection enables a much more robust and accurate identification of PDAC. Additionally, as liquid-based testing is not available for PDAC (except for CA19-9, which has significant limitations), ML-aided metabolic PDAC detection, which can distinguish benign pancreas diseases from PDAC, could benefit the detection of PDAC.

\subsection{Morphological and molecular breast cancer profiling through explainable machine learning \citep{RN159}}
The primary goal of this work was to develop a machine-learning algorithm for analyzing a variety of molecular characteristics associated with tumor-infiltrating lymphocytes (TiLs) and histological images in order to classify breast cancer based on molecular feature identification and ranking in relation to statistical associations with tumor pathology. The training and validation datasets came from TCGA.

Tumor grade and survival were determined using molecular characteristics such as protein (PROT), gene expression (RNASQ), somatic mutations (SOM), copy number variations (CNV), gene methylation (METH), and clinicopathological factors. Machine learning methods are divided into two parts: the first involves molecular characteristics, and the second involves spatial heat mapping. The heatmap was created using layer-wise relevance propagation (LRP), which is an optimization technique. Four separate application issues were addressed using support vector (SV) kernels optimized using quadratic optimization. The correlation of molecular and morphological traits with spatial predictions allows for the performance of correlation to be done, which results in an improvement in prediction performance. Because the combination of genetic and morphological data into geographical predictions allows for correlation, the performance of spatial predictions is improved even more.

However, as predictions do not necessarily indicate the physical location of the feature, the approach cannot wholly replace immunohistochemistry staining.

The model achieved over 90\% accuracy, and overlaying the original image information into the classification results is possible.

\subsection{Predicting and characterizing a cancer dependency map of tumors with deep learning \citep{RN144}}

The paper created a deep learning model, "DeepDEP," with an unsupervised pre-training structure to study the genomic profiles of tumors and predict their dependencies, which may allow them to proliferate in a malicious manner. The pan-cancer synthetic dependency map of 8,000 tumors was constructed using DeepDEP.

DeepDEP is a deep learning model with a transfer-learning and unsupervised pre-training architecture designed to analyze unlabeled cancer cell line (CCL) data for genomic associations with cancer gene dependencies. The model is divided into two primary components: encoder and predictor. The encoder neural networks are composed of five components trained with genomic profiles from TCGA and CRISPR-Cas9-based dependency profiles. Each of which encodes a distinct set of inputs, including DNA mutation, gene expression, DNA methylation, copy number alteration, and fingerprints. The fingerprint network is used to isolate fingerprints from gene dependency of interest (DepOI). The prediction network is utilized to translate learned features into dependency scores, therefore predicting the gene's dependency based on the involvement of fingerprint vectors from DepOI.

\begin{figure}[H]
\includegraphics[width=\linewidth]{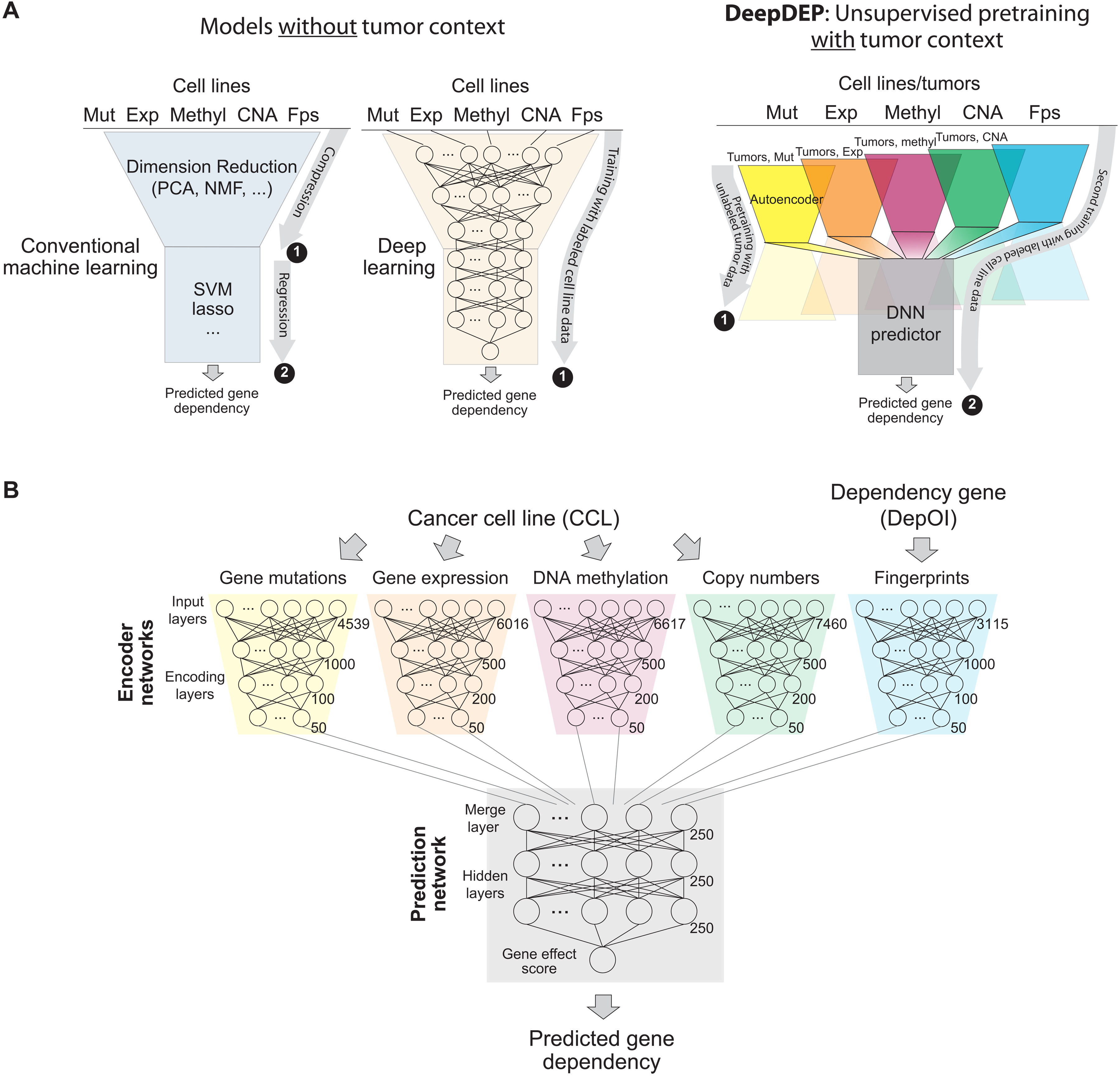}
\caption {Architecture of DeepDEP. \textbf{(A)} Pre-training study designs of conventional methods and the proposed model.
Conventional ML techniques (left) were constructed using just labeled CCL data on relevant issues like medication sensitivity prediction. A similar method may be used to implement DL models (middle). The DeepDEP model (right) gathers unlabeled tumor genomic representations and is subsequently trained using CCL gene dependency data (labeled data). Fps, functional fingerprint of gene relationships; DNN, deep neural network. \textbf{(B)} DeepDEP's design. DeepDEP predicts the gene impact score of a cancer sample's dependence of interest (DepOI) (CCL or tumor). The former was transferred from an unsupervised pretraining of autoencoders utilizing TCGA tumors. The model was trained and evaluated with Broad DepMap CCL data. \\
Note: From "Predicting and characterizing a cancer dependency map of tumors with deep learning," by \cite{RN144}, \textit{Science Advances}, 7(34). Copyright 2021 by Chiu et al. under CC BY-NC license. Reprinted under license terms.}
\label{fig:}
\end{figure}

The limitation was inadequate diversity, sample size, and intercorrelation levels of cancer cell lines may result in deviated results (seen in a standalone test of Exp-DeepDEP where correlation can't be found resulted in abundant zero values.)

The paper discovered that DeepDEP found significantly more features associated with DepOI when compared with conventional ML models and that DeepDEP can be optimized by additionally training with labeled cell line data.

\subsection{Predictions of cervical cancer identification by photonic method combined with machine learning \citep{RN158}}

In this study, the prediction algorithm was trained using low-coherence measurements of refractive index liquids extracted from cervical tissues and neoplastic lesions to classify cervical intraepithelial neoplasia (CIN) based on different stages.

As the classifier, Random Forest (RF), eXtreme Gradient Boosting (XGBoost), Naïve Bayes (NB), and Convolutional Neural Networks (CNN) were used. Filtration, enrichment, and extraction, of the refractive index data of test liquids, were carried out using the data from the Fabry-Perot interferometer. The multidimensional datasets are inputted into the supervised machine learning algorithm to classify them for the CIN. All classifiers obtained optimal results with precision, accuracy, recall, and F1 above 95\% for training datasets and around 90\% for validation sets. All classifiers achieved ideal performance, with precision, accuracy, recall, and F1 values over 95\% for training datasets and approximately 90\% for validation datasets. XGBoost and Naïve Bayes performed best in the testing set and validation set, respectively.

\begin{figure}[H]
\includegraphics[width=\linewidth]{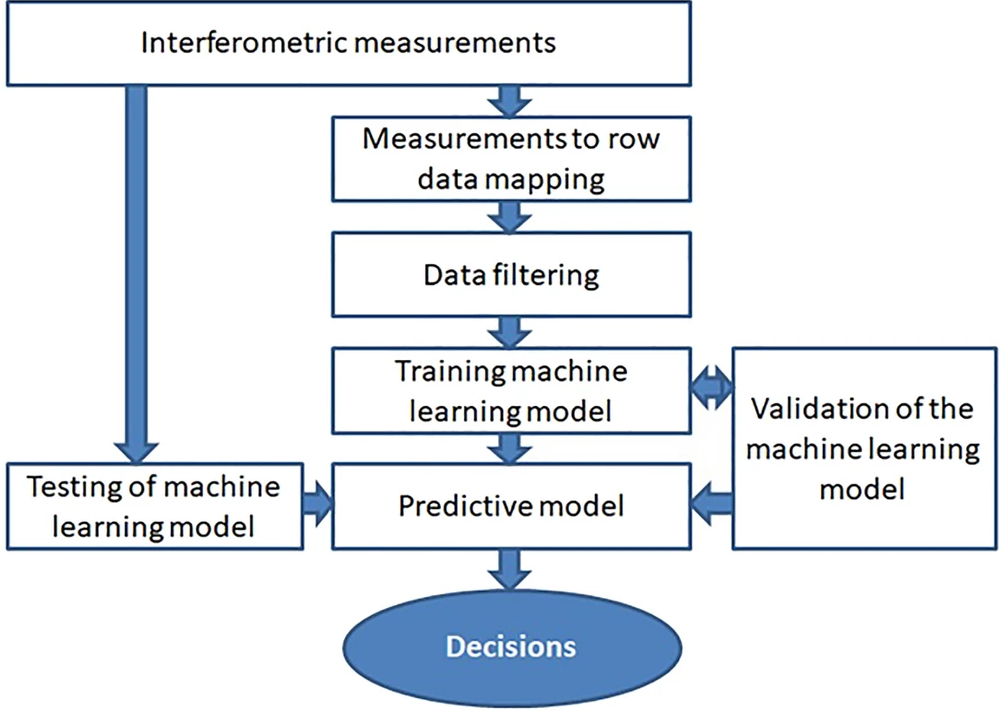}
\caption {Methodology workflow\\
Note: From "Predictions of cervical cancer identification by photonic method combined with machine learning," by \cite{RN158}, \textit{Scientific Reports}, 12, 3762. Copyright 2022 by Kruczkowski et al. under CC BY license. Reprinted under license terms.}
\label{fig:}
\end{figure}

However, extensive model overfitting was detected as a result of the learning process being excessively intensive. 
\subsection{Toward robust mammography-based models for breast cancer risk \citep{RN151}}
The study built a deep learning model termed "Mirai" that assists in predicting breast cancer risk based on mammography. It can assess several time points and identify whether or not critical risk factor information was missing during the risk assessment validation process. The approach is dedicated to improving breast cancer early detection and minimizing collateral damage during evaluation.

Mirai comprises four modules: an image encoder, an image aggregator, a risk factor predictor, and an additive-hazard layer trained on a dataset of over 80,000 patients' evaluations. Data from typical mammograms such as craniocaudal (CC) and mediolateral-oblique (MLO) images may be incorporated into the model. Each independently encoded mammogram that passes through the image aggregation module is aggregated to provide a comprehensive set of information for all available perspectives. The Tyrer-Cuzick model is then used to predict risk factors using the aggregated components. Predicted values will be used in the absence of an adequate risk factor. The additive-hazard layer incorporates risk factor data, whether predicted or provided, and aggregation elements, which are then applied to predict a patient's likelihood of developing breast cancer in the next five years.

\begin{figure}[H]
\includegraphics[width=\linewidth]{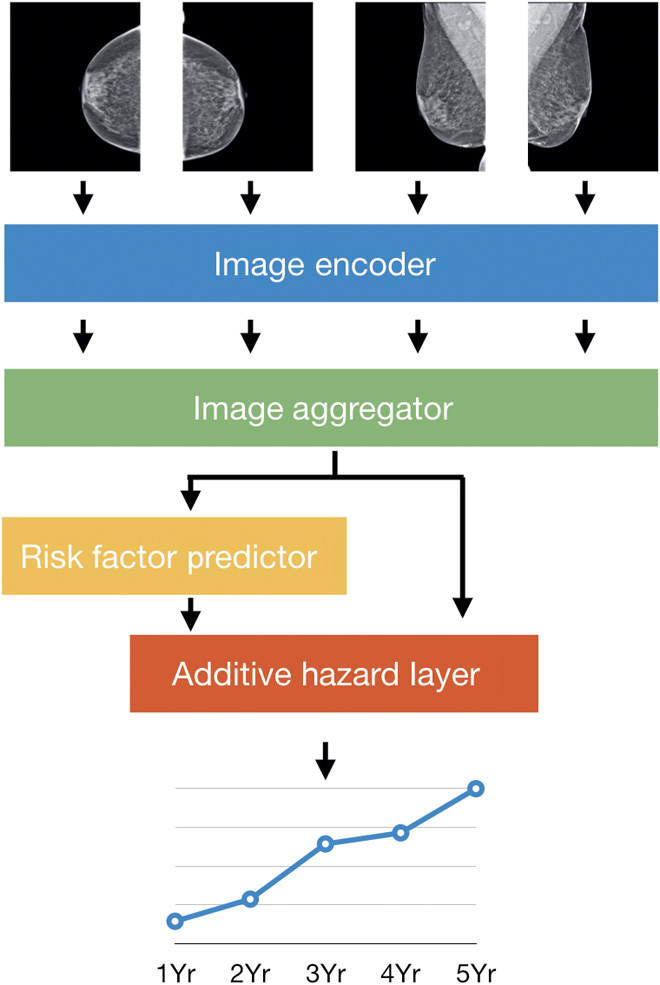}
\caption {Schematic description of Mirai. Mirai was fed four typical mammography views.
Encoding each view as a separate vector, the image aggregator aggregated all four views into a single mammography vector.
They utilized a transformer as main image aggregator and a single shared ResNet-18 as the image encoder. The risk factor prediction module projected age, full family history, and hormonal variables from the mammography vector. In the additive hazard layer, data from the image aggregator and risk variables (predicted or provided) were merged to predict risk over 5 years (Yr).\\
Note: From "Toward robust mammography-based models for breast cancer risk," by \cite{RN151}, \textit{Science Translational Medicine}, 13(578). Copyright 2021 by Yala et al. Reprinted with permission.}
\label{fig:}
\end{figure}

The limitation was that the data from several races and ethnicities is still insufficient to assess performance across races. Thus, more performance validation study is still needed before widespread employment.

It was found that Mirai greatly enhanced the true-positive rate, as determined by receiver operating characteristic curves with improved detection more noticeable in the first few years than in subsequent years.

\subsection{Tumor-specific cytolytic CD4 T cells mediate immunity against human cancer \citep{RN152}}
The study developed a deep learning network based on a deep convolutional network called DeepLabCut, which analyzes cell interactions on plates and determines the status of tumor cells and CD4 cells (whether lysed or alive). A neural network is utilized to conduct integrated phenotypic and functional characterization on a high-throughput nano biochip in order to identify tumor-specific CD4 T cells capable of mediating an immune response for cancer immunotherapies.

A modified version of DeepLabCut v2.2 was trained with six features: well lower-left corner, well upper right corner, alive tumor cell, lysed tumor cell, alive T cell, and lysed T cell. The large image data from picowell array-based time-lapse microscopy was imported into DeepLabCut to discriminate the number of cells within the wells, as well as their coordinates and status, and to analyze the specificity of target cell lysis induced by antigen-specific CD4 T cells. The mean fluorescence intensity (MFI) of lysis was then used to calculate the duration of lysis, the kinetics of lysis, and the percentage of lysis.

\begin{figure}[H]
\includegraphics[width=\linewidth]{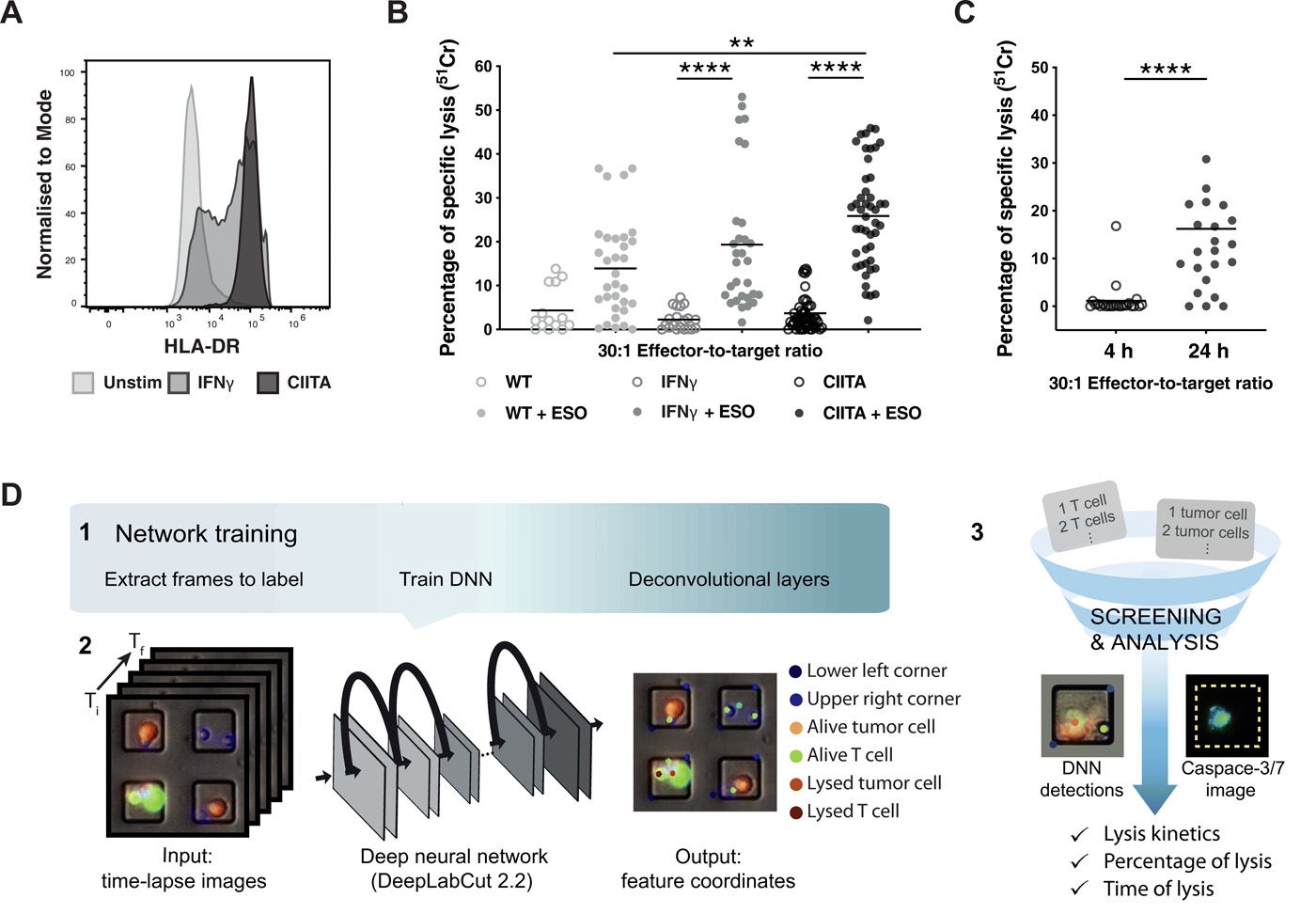}
\caption {Schematic description of Mirai. Mirai was fed four typical mammography views.
Encoding each view as a separate vector, the image aggregator aggregated all four views into a single mammography vector.
They utilized a transformer as main image aggregator and a single shared ResNet-18 as the image encoder. The risk factor prediction module projected age, full family history, and hormonal variables from the mammography vector. In the additive hazard layer, data from the image aggregator and risk variables (predicted or provided) were merged to predict risk over 5 years (Yr).\\
Note: From "Tumor-specific cytolytic CD4 T cells mediate immunity against human cancer," by \cite{RN152}, \textit{Science Advances}, 7(9). Copyright 2021 by Cachot et al. under CC BY-NC license. Reprinted under license terms.}
\label{fig:}
\end{figure}

The limitation was that comparisons of cell cytolytic capacities are difficult due to the restricted functional assessment of cytolytic ability. 

In this paper, with the help of machine learning, SLAMF7 protein was found to enhance cytotoxicity in CD4 T cells.

\section{Pathophysiology}

\subsection{A machine learning approach for somatic mutation discovery \citep{RN149}}
The research established a machine-learning-based technique called “Cerebro” for identifying tumor alterations that outperformed conventional experimental validation methods. The technique increases the possibility of excluding erroneous alterations while maximizing the sensitivity and specificity of detection for true alterations in TCGA (The Cancer Genome Atlas Programme) data.

Cerebro examines a large number of decision trees and assigns a confidence score to each variant. The model is trained on two exome genomes sequenced twice using next-generation sequencing from normal peripheral blood DNA. Over 30,000 somatic mutations were introduced into the NGS data to train the classifier. The model was exposed to over 2 million NGS errors and artifacts to ensure that errors were not mislabeled as variations. An NGS data set from the same sample was used to detect the variation. The dual alignment protocol is used to compare tumor-normal whole-exome sequences to the human reference genome to look for candidate mutations. Cerebro was used to analyze candidate mutations and provide a confidence score based on distinct coverage, MAF, GC content, and mapping quality to identify high-confidence somatic mutations.

\begin{figure}[H]
\includegraphics[width=\linewidth]{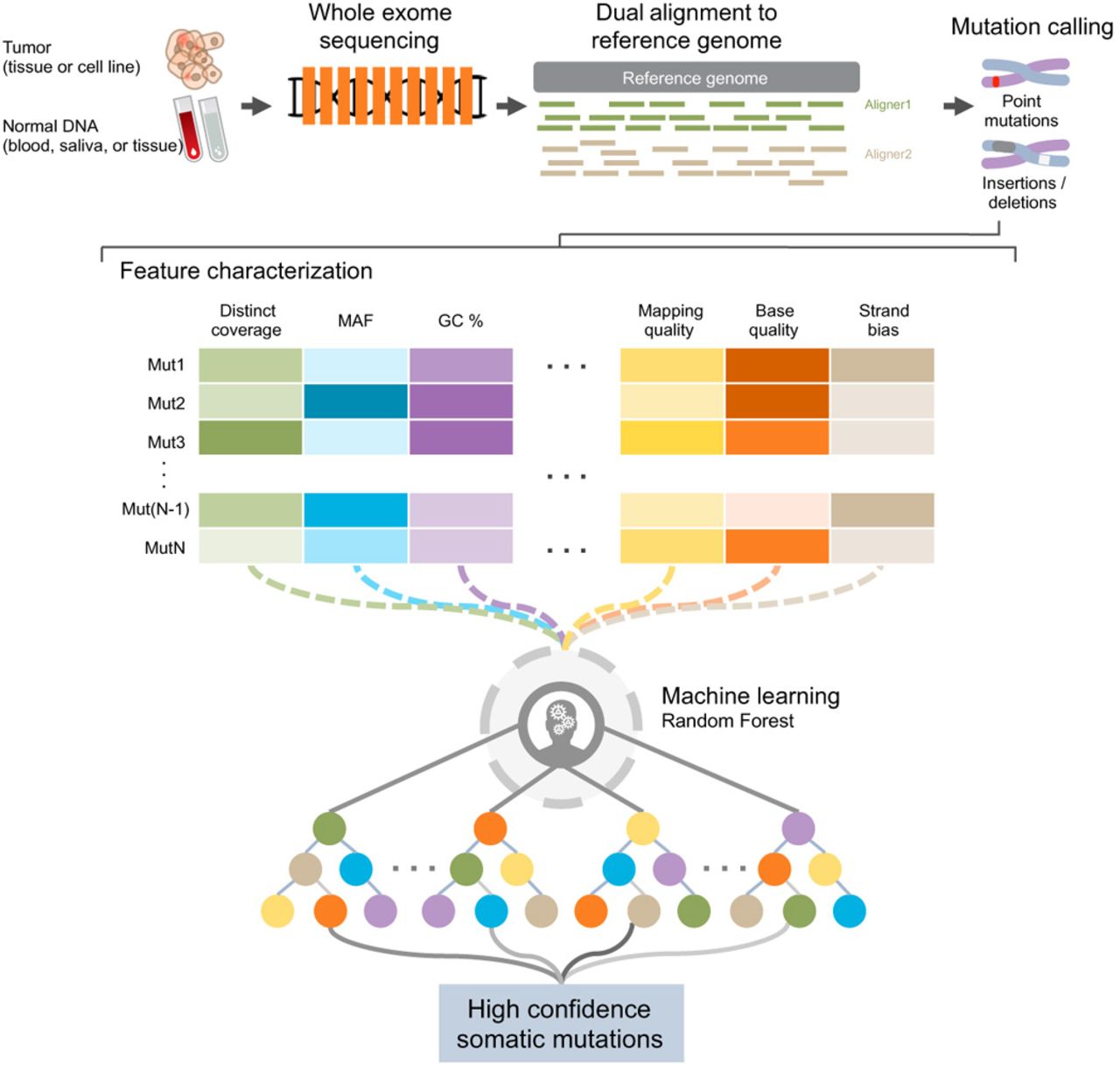}
\caption {Cerebro overview for detecting somatic mutations.  \\
Note: From "A machine learning approach for somatic mutation discovery," by \cite{RN149}, \textit{Science Translational Medicine}, 10(457). Copyright 2018 by Wood et al. Reprinted with permission.}
\label{fig:}
\end{figure}

In this paper, the training data is intended for use in conjunction with certain genomic analyses. Different data analysis coverages, such as different sequencing methods, may influence performance. Therefore, expanded training is mandatory.

Cerebro achieved a detection sensitivity of 97\% and a positive predictive value of 98\%, which outperformed all other detection methods in both positive predictive value and sensitivity. Additionally, about 10\% of the total mutations identified are uniquely called by Cerebro.

\subsection{Fusion of fully integrated analog machine learning classifier with electronic medical records for real-time prediction of sepsis onset \citep{RN160}}

The research created an on-chip fusion artificial intelligence (AI) system that combines electrocardiogram (ECG) and electronic medical records (EMR) data to predict sepsis onset based on a score. The physiological sensor transmits data to the cloud AI model, which is then analyzed along with EMR, circumventing the constraints of most wearable devices, such as the Apple Watch, which has yet to incorporate EMR in real-time.

On the chip, ECG is preprocessed, and feature extraction was accomplished by determining the signal median value and extracting features using first-order statistics of R-R peaks. For analysis, central tendency, dispersion, the shape of the distribution of a window, R-R peaks, R-R intervals, variance, and average heart rate are extracted for analysis. The Artificial Neural Network (ANN) architectures trained with the training dataset were then incorporated. Along with the ECG, the EMR model analyzes patients' co-morbidities using vectorized ICD10 codes and the Term Frequency-Inverse Document Frequency Algorithm (TF-IDF) to predict the probability of sepsis. The two models generate the sepsis score and aggregate it into a fusion model that the meta-learner (linear support vector machine, logistic regression, random forest, and neural network) analyzes for the final prediction of sepsis risk.

\begin{figure}[H]
\includegraphics[width=\linewidth]{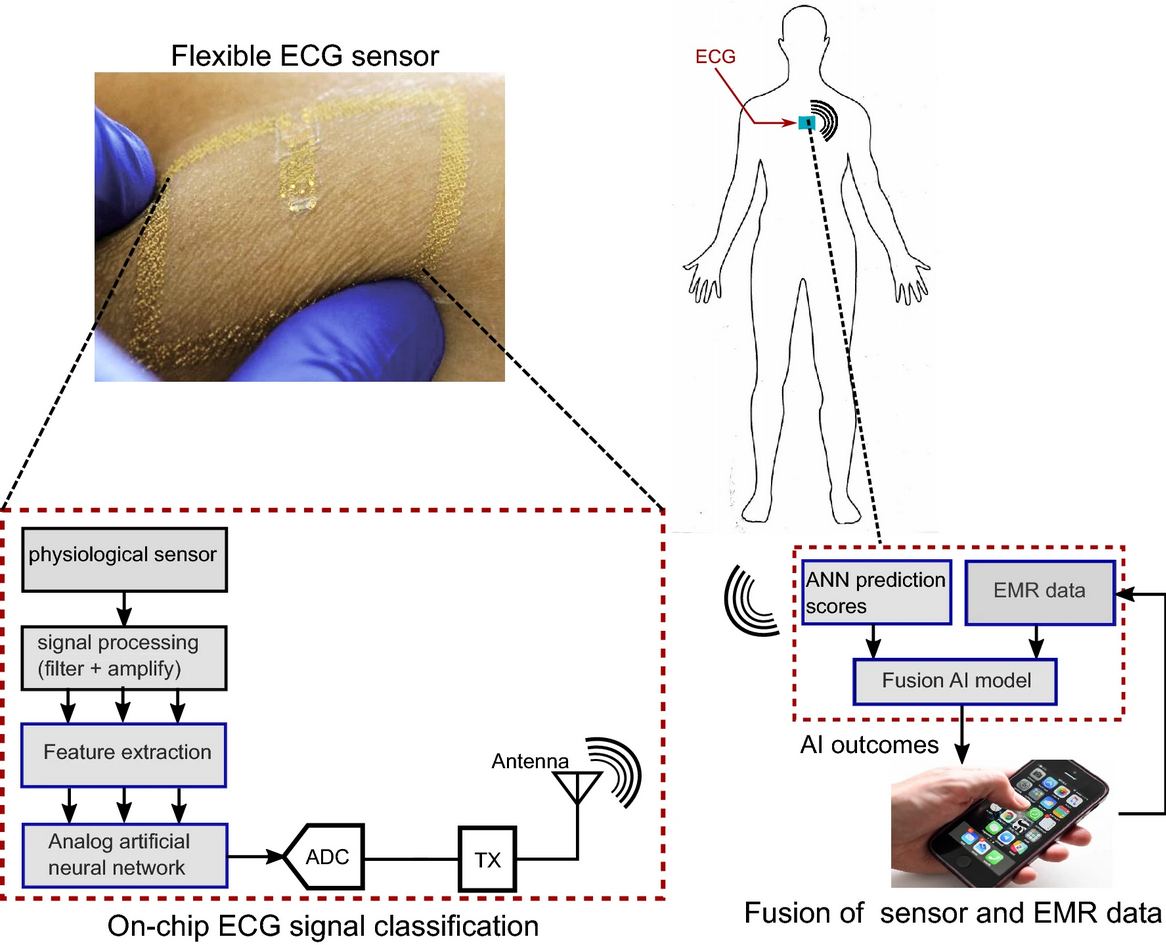}
\caption {A diagrammatic representation of the proposed model, with components highlighted in blue indicating the tasks addressed in this study.\\
Note: From "Fusion of fully integrated analog machine learning classifier with electronic medical records for real-time prediction of sepsis onset," by \cite{RN160}, \textit{Scientific Reports}, 12, 5711. Copyright 2022 by Sadasivuni et al. under CC BY license. Reprinted under license terms.}
\label{fig:}
\end{figure}

In this study, the AI model (neural network layer, feature extractor, preprocessor, etc.) was hardcoded into the chip, thus, preventing the chip from being updated in the future.

It was concluded that the chip and fusion model could predict sepsis 4 hours before the onset at 93\% accuracy. 

\subsection{The human splicing code reveals new insights into the genetic determinants of disease \citep{RN148}}

This work established a computational algorithm to determine the RNA splicing's dependence on distinct genetic variations. The model may be used to predict human tissue splicing and to investigate the consequences of common, rare, and spontaneous DNA variations that may cause disease. The finding demonstrates that the model is capable of classifying disease-causing variations.

A Bayesian deep learning approach is used to create the computational model. The splicing regulation model may be applied to any sequence with a triplet of exons. The model was trained using 1,393 sequence features collected from each exon and its adjacent introns and exons retrieved from 10,689 exons with evidence of alternative splicing. The model extracts DNA sequence features from a particular cell type and compares them to normal human splicing levels in order to predict the percentage of transcripts with the central exon spliced in (Ψ) and the Bayesian confidence estimate. Genetic variations caused by diseases can be filtered and found using the variant score in the computationally estimated splicing levels.

\begin{figure}[H]
\includegraphics[width=\linewidth]{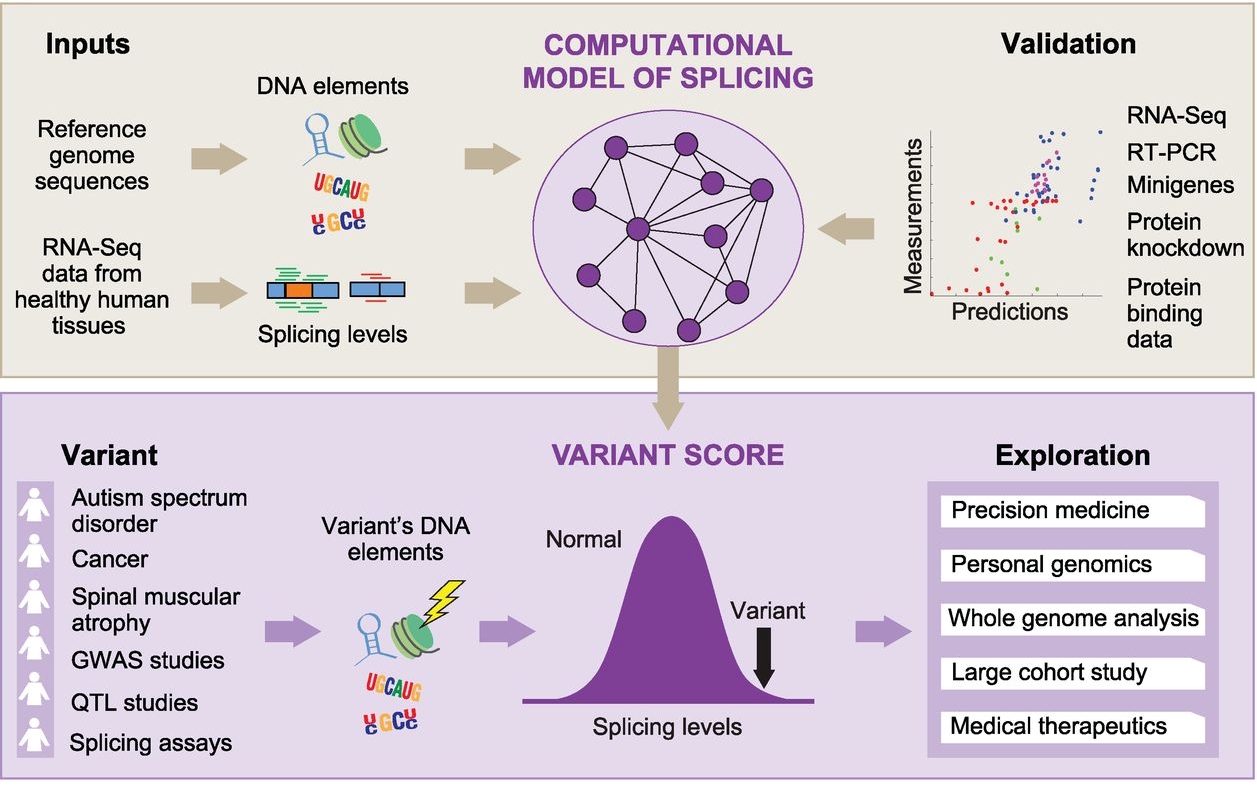}
\caption {Detecting damaging genetic variations using a splicing model. \textbf{(A)} Top: Correlating DNA elements with splicing levels in healthy human tissues helps infer a computer model of splicing. Bottom: The computer model can identify and filter genetic variations from a variety of illnesses and technologies, allowing for disease genetic research.\\
Note: From "The human splicing code reveals new insights into the genetic determinants of disease," by \cite{RN148}, \textit{Science Advances}, 7(47). Copyright 2015 by American Association for the Advancement of Science. Reprinted with permission.}
\label{fig:}
\end{figure}

The limitation was that unaccounted-for RNA characteristics, inaccuracies in computed features, and imprecise modeling of splicing levels might lead to prediction errors.

This paper found that 94\% of variants associated with altered splicing by minigene reporters were correctly classified and evaluation of more than 650,000 variants revealed that disease-causing variants have a higher score with disease annotated introns were found to have higher regulatory scores than GWAS and non-GWAS.

\section{Pharmaceutical}

\subsection{A machine learning approach to define antimalarial drug action from heterogeneous cell-based screens \citep{RN145}}

The study developed a semi-supervised machine learning architecture to assess malaria parasite cultures and classify their development stages according to their morphological heterogeneity. Diverse morphologies of Plasmodium falciparum were defined and quantified throughout the asexual life cycle using the model to characterize its development and antimalarial medication effects.

Numerous human labels for malaria's asexual development were obtained to be utilized in conjunction with the unlabeled data. Unlabeled malaria photos were turned into input embeddings using a deep metric network, resulting in a 384-dimensional embedding for each image. The embedding from the unlabeled data was fed into a semi-supervised Deep Neural Network (DNN) (restricted to Dimethyl Sulfoxide controls) to produce pseudo labels for the unlabeled data. To get life cycle and phenotypic data, pseudo-labeled and human-labeled life cycle image data and compound image data are fed into the second semi-supervised DNN (the same DNN is used to quantify drug effects.)

\vspace{-2cm}
\begin{figure}[H]
\includegraphics[width=\linewidth]{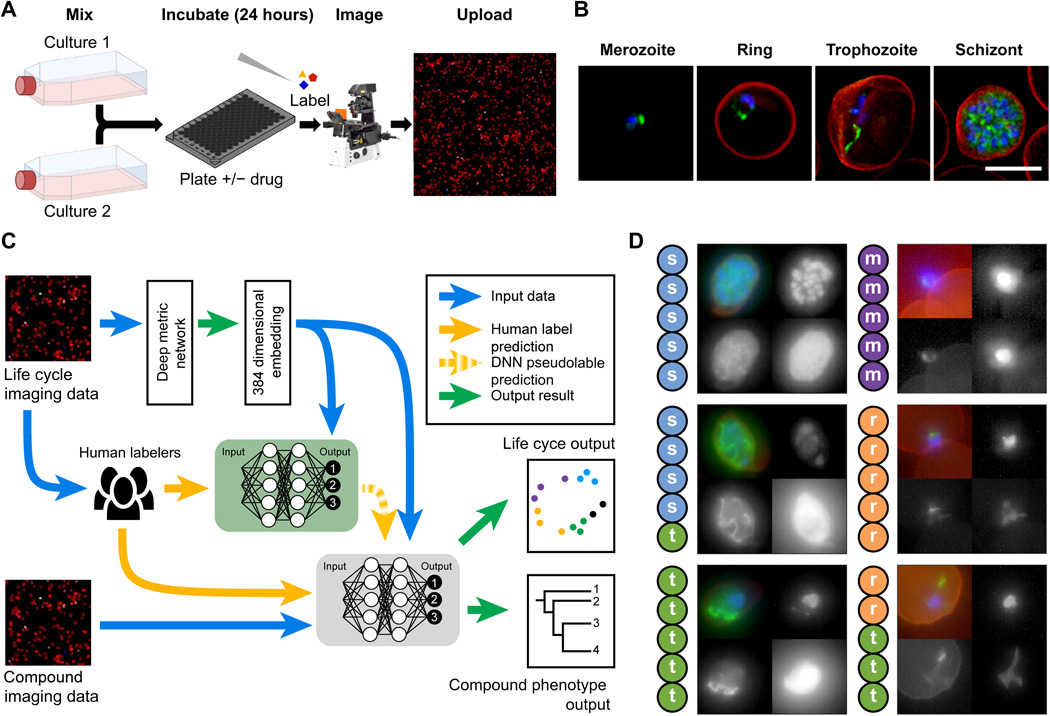}
\caption {Experimental workflow. \textbf{(A)} To ensure all life cycle phases were captured during imaging and analysis, two transgenic malaria cultures expressing sfGFP were merged. \textbf{(B)} parasite nuclei (blue), cytoplasm (not depicted), mitochondrion (green), RBC plasma membrane (red), and brightfield (not shown). The merozoite, ring, trophozoite, and schizont phases are depicted here. These images were ML analyzed. \textbf{(C)} Utilizing the nuclear stain channel to separate parasites from full field images before embedding vectors. The first semi-supervised network (green) predicted life cycle stage and compound phenotype using ML–derived pseudolabels obtained from human-labeled imaging data. Human-labeled datasets examples \textbf{(D)} illustrate that human labelers may differ on parasite stage classification (s, schizont; t, trophozoite; r, ring; m, merozoite). Each thumbnail picture displays merged channels, nucleus staining, cytoplasm, and mitochondria (clockwise). 5 \micro m scale bar.\\
Note: From "A machine learning approach to define antimalarial drug action from heterogeneous cell-based screens," by \cite{RN145}, \textit{Science Advances}, 6(39). Copyright 2020 by Ashdown et al. under CC BY license. Reprinted under license terms.}
\label{fig:}
\end{figure}

The limitation was that deep neural network models and human-supervised data are required, and inconsistencies may arise between labelers in multi-labeler settings.

This paper found that outliers of the parasite not seen during asexual development can be detected by the DNN, and based on canonical human-labeled data, the DNN properly organized the continuum of parasite stages visualized the on-cycle drug effects.

\subsection{Combining generative artificial intelligence and on-chip synthesis for de novo drug design \citep{RN142}}
This study looked at a design-make-test-analyze (DMTA) cycle that used deep learning with an on-chip chemical synthesis microfluidics technology to generate LXR agonists for the liver. The screening demonstrates that the LXRs produced by our AI-based on-chip synthesis are powerful LXR agonists, indicating a possible new therapeutic design that can be investigated further.

A generative deep learning model based on a recurrent neural network with long-short term memory (LSTM) cells functioning as a de novo structure generator was pre-trained using 656,070 molecules in order to choose compounds appropriate for the microfluidics system. A fine-tuning method for the generator's model was then carried out using 40 known LXR$\alpha$ agonists. It is then utilized to create a de novo drug structure for LXR agonists, which is subsequently tested. Following the generation of potential LXR agonists using the generative deep learning model, the outputs are filtered by 17 sets of virtual response filters in order to ensure compatibility with automated on-chip synthesis, as described before. Following the completion of the on-chip synthesis, the final sample was evaluated using high-performance liquid chromatography-mass spectrometry (HPLC-MS) prior to collection.

\begin{figure}[H]
\includegraphics[width=\linewidth]{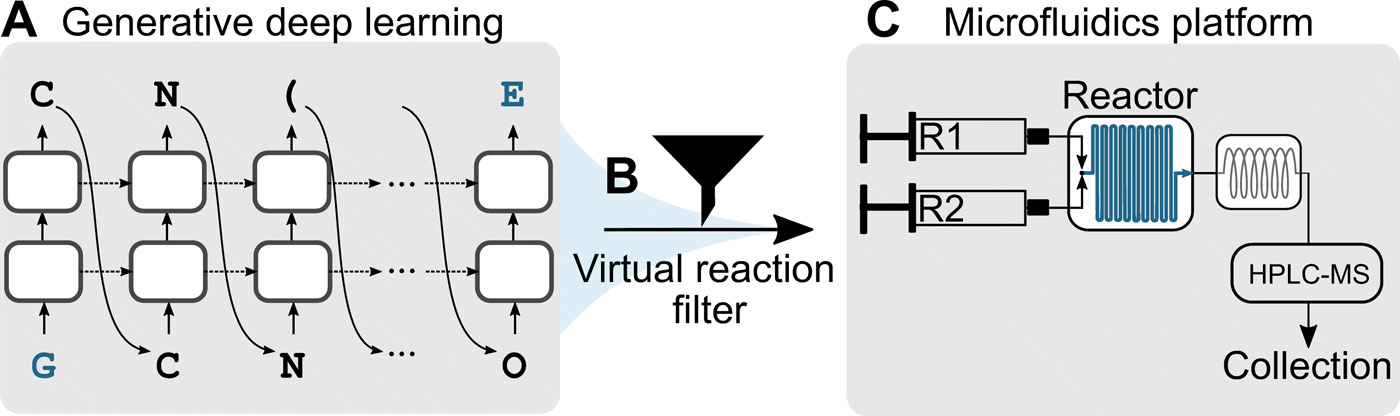}
\caption {The modular molecular design process is shown schematically. \textbf{(A)} For the liver X receptors, we employed a deep learning model with a long-short term memory network. First, the network was trained on 656,070 chemicals anticipated to be compatible with the microfluidics technique, and then fine-tuned using 40 known LXR$\alpha$ agonists. \textbf{(B)} The de novo produced compounds were screened using a set of 17 SMARTS reactions (“virtual reaction filter”). Synthetic molecular building components were automatically acquired from a supplier list. \textbf{(C)} On the microfluidics platform, 41 de novo designs were chosen to be synthesized using two syringes for reagents R1 and R2, a Cetoni Qmix element with a Dean Flow microfluidic reactor chip, and a Rheodyne MRA splitter for automated sample transfer to an HPLC-MS system.\\
Note: From "Combining generative artificial intelligence and on-chip synthesis for de novo drug design," by \cite{RN142}, \textit{Science Advances}, 7(24). Copyright 2021 by Grisoni et al. under CC BY-NC license. Reprinted under license terms.}
\label{fig:}
\end{figure}

Note that in this paper, the activation validation has only been carried in vitro only. In vivo activation has not been tested.

In this paper, it was found that 55\% of de novo designs generated by the generative deep learning model were not found in either PubChem, ChEMBL27, SciFinder, SureChEMBL, or Reaxys database, suggesting they are novel and fragment similarity of the de novo designs were observed to be like their fine-tuning compound.

\subsection{Supervised learning model predicts protein adsorption to carbon nanotubes \citep{RN146}}

The research studied and created a supervised random forest classifier (RFC) that could classify proteins with high binding affinity to a single-walled carbon nanotube (SWCNT) purely based on the sequence of the protein with high precision and accuracy. Based on the previous work, the new model is refined with improved threshold settings associated with the corona phase.

The Protein corona dataset was developed using a selective absorption method based on LC-MS/MS characterization. Additionally, an array of anticipated physicochemical protein attributes was constructed using amino acid sequences from UniProt, an annotated protein database. Additional biological protein features were also included (however, testing suggested only marginal performance improvement.) Two criteria were used to establish a threshold for protein placement in or out of the corona: relative change and abundance threshold. RFC was shown to be the most effective method for classifying corona proteins on (GT)15-SWCNT nanoparticles. Stratified shuffle split validation was used to determine the classifier's accuracy, with the dataset partitioned into training and test sets. The training set is used to train an untrained classifier, which is then assessed by a test classifier and compared to true responses in an iterative process. ANOVA (analysis of variance) was used to compare protein properties that may be significant for the categorization of nanoparticle corona to the predictions of a classifier based on distributions. The model was assessed and validated by being used to predict the interaction affinity of more than 2,000 total proteins. Protein binding dynamics were compared with a real-time corona exchange assay.

\begin{figure}[H]
\includegraphics[width=\linewidth]{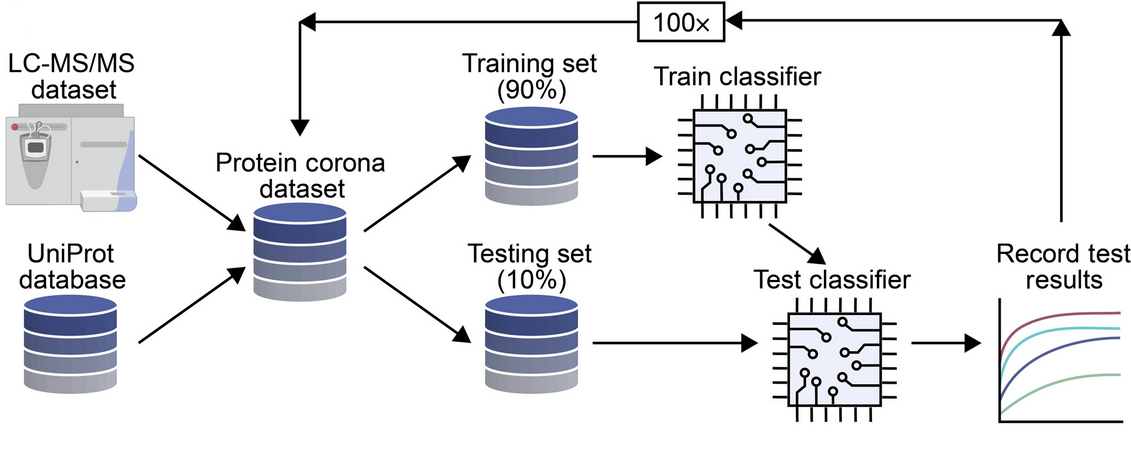}
\caption {RFC workflow for splitting-based predictions. LC-MS/MS gives protein corona composition experimentally. To create a comprehensive dataset, LC-MS/MS data are merged with protein attributes extracted from protein sequences (UniProt database using BioPython module). 90\% of the dataset is for training and 10\% for testing. Training data are used to grade a reset classifier, and the cycle starts again. \\
Note: From "Supervised learning model predicts protein adsorption to carbon nanotubes," by \cite{RN146}, \textit{Science Advances}, 8(1). Copyright 2022 by Ouassil et al. under CC BY license. Reprinted under license terms.}
\label{fig:}
\end{figure}

As protein corona is still a poorly understood phenomenon, nanoparticle-based technologies are thus limited. Additionally, several nanotechnologies, including SWCNTs, often exhibit unanticipated interactions, further complicating their design. The discrepancy also arises with the inclusion of proteins that may be present on the SWCNTs from one specific biofluid but not present on the SWCNTs from another biofluid.

This paper found that the supervised learning approach with the random forest classifier can predict protein absorption efficiently.

\subsection{Systems biology and machine learning approaches identify drug targets in diabetic nephropathy \citep{RN155}}
In this study, a machine learning algorithm based on "modified Group Method of Data Handling with Automatic Feature Selection (mGMDH-AFS)" capable of tolerating class size imbalance was developed to classify novel drug targets (DTs) for diabetic nephropathy (DN). The model significantly outperformed the current technology used to identify potential DTs.

\begin{figure}[H]
\includegraphics[width=\linewidth]{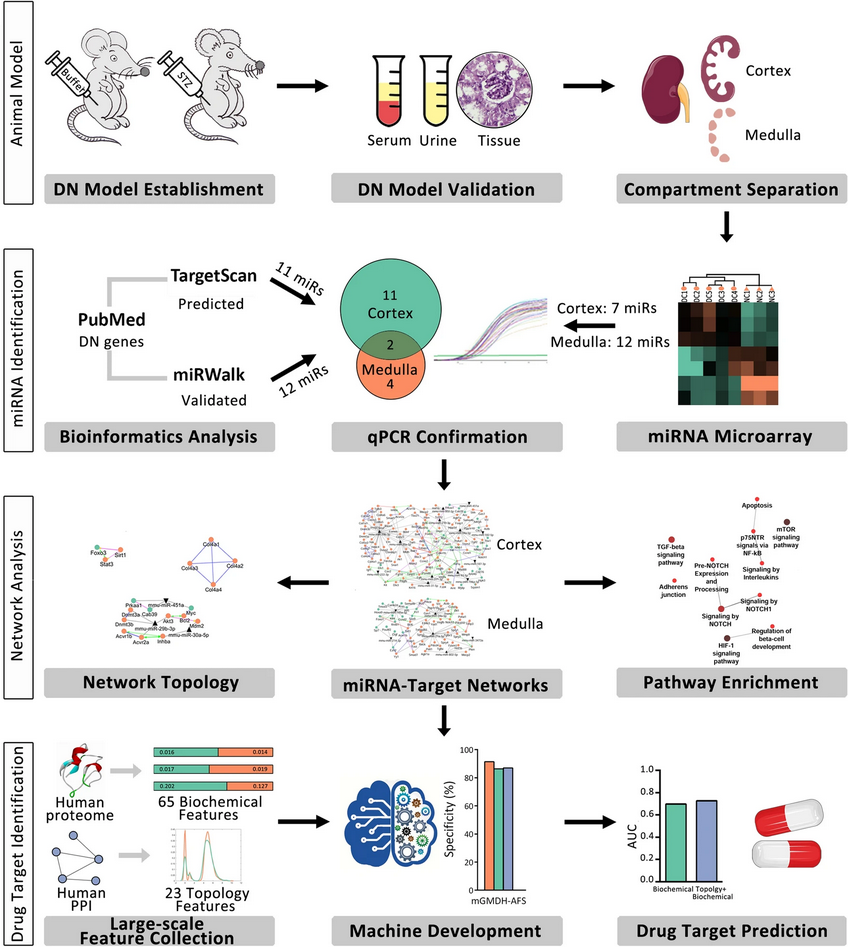}
\caption {A study design schematic. This study's aim was to predict new therapeutic targets for DN using a molecular pathogenesis map. The major nodes, essential connections, and signaling pathways of DN were found experimentally and computationally. mGMDH-AFS, a high-performance machine learning framework, was designed and proven to predict drug targets for all human proteins. This classifier was then used to fresh therapeutic target candidates in the DN holistic map. miRs: microRNAs; PPI: protein–protein interaction network. \\
Note: From "Systems biology and machine learning approaches identify drug targets in diabetic nephropathy," by \cite{RN155}, \textit{Scientific Reports}, 11, 23452. Copyright 2021 by Abedi et al. under CC BY license. Reprinted under license terms.}
\label{fig:}
\end{figure}

The identification of miRNA profiles was performed in order to identify differentially expressed (DE) miRNAs in diabetic kidneys. The quantitative polymerase chain reaction (qPCR) is used to confirm the interaction of DN genes with predicted, identified, or validated miRNAs. Pathway enrichment analysis predicts signaling pathways associated with the disorder. Then, utilizing biochemical and network topology data, mGMDH-AFS was applied to predict DT in the human proteome. Due to the distinction between DT and non-DT proteins, the cortex and medulla networks analyzed by mGMDH-AFS is capable of identifying a possible de novo drug related to DT proteins. The model is able to discover a potential de novo drug associated with DT proteins.

The constraint was that there is only a limited amount of experimental data available for miRNA profiling. Additionally, the genomics layer that is network-integrated is constrained. 

In this study, the model performed at 90\% sensitivity, 86\% specificity, 88\% accuracy, and 89\% precision.

\section{Therapeutic}

\subsection{Diving beetle–like miniaturized plungers with reversible, rapid biofluid capturing for machine learning–based care of skin disease \citep{RN150}}

\begin{figure}[H]
\includegraphics[width=\linewidth]{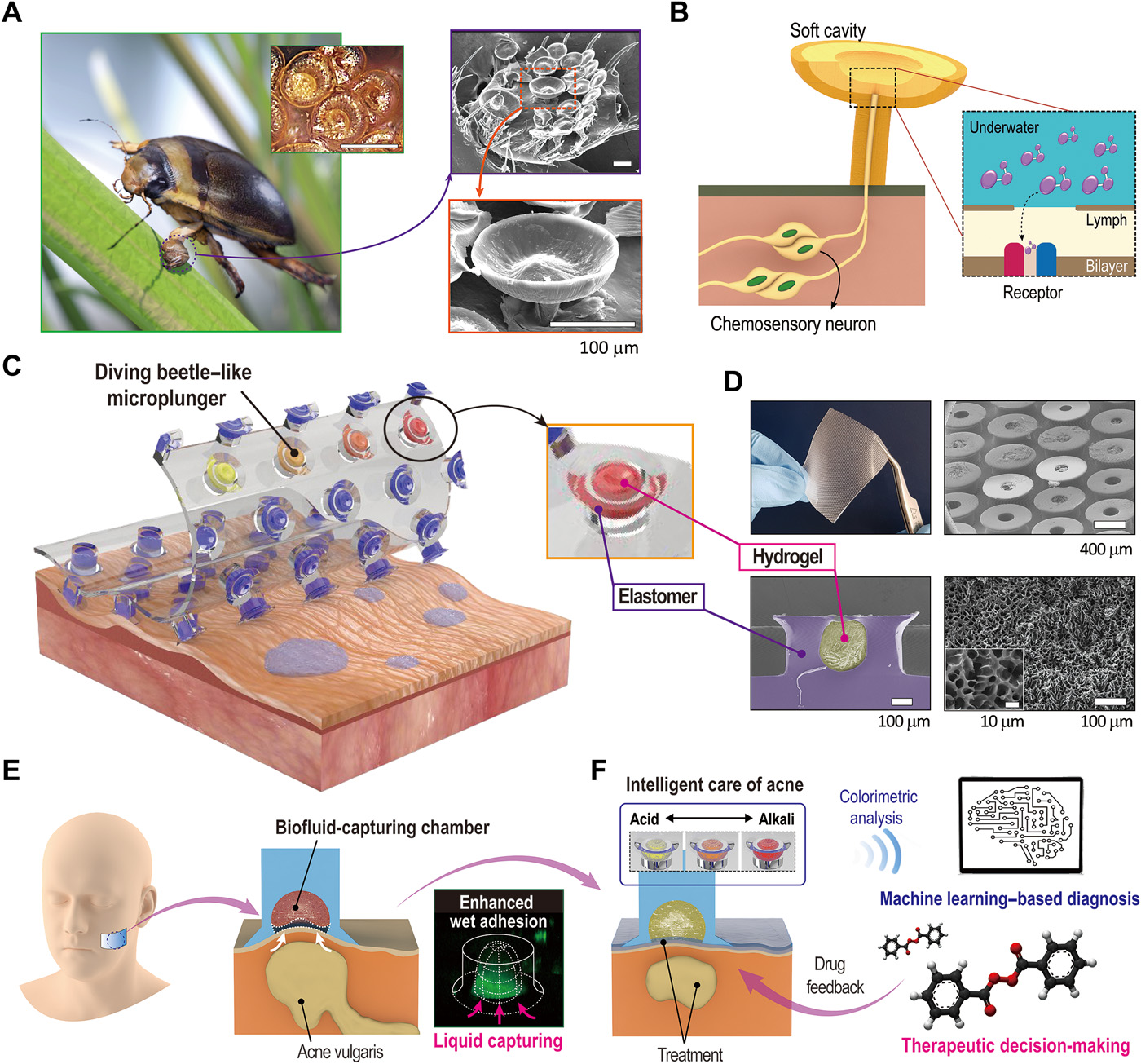}
\caption {The reversible microplungers with biofluid-capturing hydrogel resemble diving beetles. \textbf{(A)} Photograph of \textit{H. pacificus} forelegs. \textbf{(B)} Ilustration of spatula setae for suction. \textbf{(C)} The adhesive patch with suction chambers inspired by diving beetles against rough and moist human skin. \textbf{(D)} Images of the adhesive patch with suction chambers integrated with biofluid-capturing hydrogels. \textbf{(E)} Suction chambers with integrated hydrogel. \textbf{(F)} Adhesive patch arranged with suction chambers for untethered skin pH analysis.\\
Note: From "Diving beetle–like miniaturized plungers with reversible, rapid biofluid capturing for machine learning–based care of skin disease," by \cite{RN150}, \textit{Science Advances}, 7(25). Copyright 2021 by Baik et al. under CC BY-NC license. Reprinted under license terms.}
\label{fig:}
\end{figure}

The research team developed and demonstrated a microplunger and hydrogel based on a male diving beetle that can adhere to human skin in several directions. Following attachment, the hydrogels contained inside the micropungers instantly absorb liquids from the epidermis, ensuring excellent adhesiveness and serving as skin pH sensors. A machine learning framework is implemented to perform an automated colorimetric analysis of pH levels accurately.

A large mushroom-shaped rigid cup shape with a circular chamber is employed in the microplunger device to enable conformal sealing while simultaneously generating a suction effect, similar to the spatula setae of the diving beetles. In the device are biofluid-capturing hydrogels, allowing the device to be used for biomarker diagnostics. A machine learning-based model for pH quantification was utilized to rapidly distinguish pH values non-invasively by utilizing input data consisting of red-green-blue (RGB) color values designated by the hydrogen in the microplunger device.

The limitation was that further investigation of skin Ph value diversity across ethnicities and genders had not been accounted for in the machine learning algorithm.

It was found that diving beetle-inspired architectures (DIAs) obtained high adhesion both in dry and underwater conditions, with hydrogel-embedded soft-tips DIA (H-s-DIAs) having more durability while retaining adhesion. A predicted pH from the usage of DIAs and actual pH values have a significant correlation.

\subsection{}

\section{Data availability}

Data sharing not applicable to this report as no datasets were generated or analysed during the current study.

\section{Funding}
This article received no specific grant from any funding agency in the public, commercial, or not-for-profit sectors.
\section{Conflicts of interest}
The author reports no conflict of interest.

\bibliography{bibliography}

\begin{thebibliography}{29}

\bibitem[{Abedi {\em et~al.\/}}(2021){Abedi, Marateb, Mohebian,
  Aghaee-Bakhtiari, Nassiri, {\rm and} Gheisari}]{RN155}
{Abedi M, Marateb HR, Mohebian MR, Aghaee-Bakhtiari SH, Nassiri SM, Gheisari
  Y}. 2021. Systems biology and machine learning approaches identify drug
  targets in diabetic nephropathy. Scientific Reports. 11:23452.

\bibitem[{Amin {\em et~al.\/}}(2019){Amin, Sharif, Raza, Saba, {\rm and}
  Anjum}]{RN139}
{Amin J, Sharif M, Raza M, Saba T, Anjum MA}. 2019. Brain tumor detection using
  statistical and machine learning method. Computer Methods and Programs in
  Biomedicine. 177:69--79.

\bibitem[{Ashdown {\em et~al.\/}}(2020){Ashdown, Dimon, Fan, Terán, Witmer,
  Gaboriau, Armstrong, Ando, {\rm and} Baum}]{RN145}
{Ashdown GW, Dimon M, Fan M, Terán FSR, Witmer K, Gaboriau DCA, Armstrong Z,
  Ando DM, Baum J}. 2020. A machine learning approach to define antimalarial
  drug action from heterogeneous cell-based screens. Science Advances.
  6:eaba9338.

\bibitem[{Baik {\em et~al.\/}}(2021){Baik, Lee, Jeon, Park, Kim, Song, Lee,
  Han, Cho, {\rm and} Pang}]{RN150}
{Baik S, Lee J, Jeon EJ, Park By, Kim DW, Song JH, Lee HJ, Han SY, Cho SW, Pang
  C}. 2021. Diving beetle-like miniaturized plungers with reversible, rapid
  biofluid capturing for machine learning-based care of skin disease. Science
  Advances. 7:eabf5695.

\bibitem[{Binder {\em et~al.\/}}(2021){Binder, Bockmayr, Hägele, Wienert,
  Heim, Hellweg, Ishii, Stenzinger, Hocke, Denkert, Müller, {\rm and}
  Klauschen}]{RN159}
{Binder A, Bockmayr M, Hägele M, Wienert S, Heim D, Hellweg K, Ishii M,
  Stenzinger A, Hocke A, Denkert C {\em et~al\/}}. 2021. Morphological and
  molecular breast cancer profiling through explainable machine learning.
  Nature Machine Intelligence. 3:355--366.

\bibitem[{Buch {\em et~al.\/}}(2018){Buch, Ahmed, {\rm and}
  Maruthappu}]{Buch143}
{Buch VH, Ahmed I, Maruthappu M}. 2018. Artificial intelligence in medicine:
  current trends and future possibilities. British Journal of General Practice.
  68:143--144.

\bibitem[{Bulten {\em et~al.\/}}(2022){Bulten, Kartasalo, Chen, Ström,
  Pinckaers, Nagpal, Cai, Steiner, van Boven, Vink, Hulsbergen-van~de Kaa,
  van~der Laak, Amin, Evans, van~der Kwast, Allan, Humphrey, Grönberg,
  Samaratunga, Delahunt, Tsuzuki, Häkkinen, Egevad, Demkin, Dane, Tan,
  Valkonen, Corrado, Peng, Mermel, Ruusuvuori, Litjens, Eklund, Brilhante,
  Çakır, Farré, Geronatsiou, Molinié, Pereira, Roy, Saile, Salles,
  Schaafsma, Tschui, Billoch-Lima, Pereira, Zhou, He, Song, Sun, Yoshihara,
  Yamaguchi, Ono, Shen, Ji, Roussel, Zhou, Chai, Weng, Grechka, Shugaev,
  Kiminya, Kovalev, Voynov, Malyshev, Lapo, Campos, Ota, Yamaoka, Fujimoto,
  Yoshioka, Juvonen, Tukiainen, Karlsson, Guo, Hsieh, Zubarev, Bukhar, Li, Li,
  Speier, Arnold, Kim, Bae, Kim, Lee, Park, {\rm and} the}]{RN157}
{Bulten W, Kartasalo K, Chen PHC, Ström P, Pinckaers H, Nagpal K, Cai Y,
  Steiner DF, van Boven H, Vink R {\em et~al\/}}. 2022. Artificial intelligence
  for diagnosis and gleason grading of prostate cancer: the panda challenge.
  Nature Medicine. 28:154--163.

\bibitem[{Cachot {\em et~al.\/}}(2021){Cachot, Bilous, Liu, Li, Saillard,
  Cenerenti, Rockinger, Wyss, Guillaume, Schmidt, Genolet, Ercolano, Protti,
  Reith, Ioannidou, Leval, Trapani, Coukos, Harari, Speiser, Mathis, Gfeller,
  Altug, Romero, {\rm and} Jandus}]{RN152}
{Cachot A, Bilous M, Liu YC, Li X, Saillard M, Cenerenti M, Rockinger GA, Wyss
  T, Guillaume P, Schmidt J {\em et~al\/}}. 2021. Tumor-specific cytolytic cd4
  t cells mediate immunity against human cancer. Science Advances. 7:eabe3348.

\bibitem[{Chiu {\em et~al.\/}}(2021){Chiu, Zheng, Wang, Iskra, Rao, Houghton,
  Huang, {\rm and} Chen}]{RN144}
{Chiu YC, Zheng S, Wang LJ, Iskra BS, Rao MK, Houghton PJ, Huang Y, Chen Y}.
  2021. Predicting and characterizing a cancer dependency map of tumors with
  deep learning. Science Advances. 7:eabh1275.

\bibitem[{Cui {\em et~al.\/}}(2022){Cui, Fang, Mei, Zhang, Yu, Liu, Jiang, Sun,
  Ma, Huang, Liu, Zhao, Lian, Ding, Zhu, {\rm and} Shen}]{RN162}
{Cui Z, Fang Y, Mei L, Zhang B, Yu B, Liu J, Jiang C, Sun Y, Ma L, Huang J {\em
  et~al\/}}. 2022. A fully automatic ai system for tooth and alveolar bone
  segmentation from cone-beam ct images. Nature Communications. 13:2096.

\bibitem[{Ganzer {\em et~al.\/}}(2022){Ganzer, Loeian, Roof, Teng, Lin,
  Friedenberg, Baumgart, Meyers, Chun, Rich, Tsao, Muir, Weber, {\rm and}
  Hamlin}]{RN136}
{Ganzer PD, Loeian MS, Roof SR, Teng B, Lin L, Friedenberg DA, Baumgart IW,
  Meyers EC, Chun KS, Rich A {\em et~al\/}}. 2022. Dynamic detection and
  reversal of myocardial ischemia using an artificially intelligent
  bioelectronic medicine. Science Advances. 8:eabj5473.

\bibitem[{Grisoni {\em et~al.\/}}(2021){Grisoni, Huisman, Button, Moret, Atz,
  Merk, {\rm and} Schneider}]{RN142}
{Grisoni F, Huisman BJH, Button AL, Moret M, Atz K, Merk D, Schneider G}. 2021.
  Combining generative artificial intelligence and on-chip synthesis for de
  novo drug design. Science Advances. 7:eabg3338.

\bibitem[{Hamet {\rm and} Tremblay}(2017)]{RN164}
{Hamet P, Tremblay J}. 2017. Artificial intelligence in medicine. Metabolism.
  69:S36--S40.

\bibitem[{Hussein {\em et~al.\/}}(2019){Hussein, Kandel, Bolan, Wallace, {\rm
  and} Bagci}]{RN140}
{Hussein S, Kandel P, Bolan CW, Wallace MB, Bagci U}. 2019. Lung and pancreatic
  tumor characterization in the deep learning era: Novel supervised and
  unsupervised learning approaches. IEEE Transactions on Medical Imaging.
  38:1777--1787.

\bibitem[{Jamshidi {\em et~al.\/}}(2020){Jamshidi, Lalbakhsh, Talla, Peroutka,
  Hadjilooei, Lalbakhsh, Jamshidi, Spada, Mirmozafari, Dehghani, Sabet,
  Roshani, Roshani, Bayat-Makou, Mohamadzade, Malek, Jamshidi, Kiani,
  Hashemi-Dezaki, {\rm and} Mohyuddin}]{RN141}
{Jamshidi M, Lalbakhsh A, Talla J, Peroutka Z, Hadjilooei F, Lalbakhsh P,
  Jamshidi M, Spada LL, Mirmozafari M, Dehghani M {\em et~al\/}}. 2020.
  Artificial intelligence and covid-19: Deep learning approaches for diagnosis
  and treatment. IEEE Access. 8:109581--109595.

\bibitem[{Jurmeister {\em et~al.\/}}(2019){Jurmeister, Bockmayr, Seegerer,
  Bockmayr, Treue, Montavon, Vollbrecht, Arnold, Teichmann, Bressem, Schüller,
  Laffert, Müller, Capper, {\rm and} Klauschen}]{RN138}
{Jurmeister P, Bockmayr M, Seegerer P, Bockmayr T, Treue D, Montavon G,
  Vollbrecht C, Arnold A, Teichmann D, Bressem K {\em et~al\/}}. 2019. Machine
  learning analysis of dna methylation profiles distinguishes primary lung
  squamous cell carcinomas from head and neck metastases. Science Translational
  Medicine. 11:eaaw8513.

\bibitem[{Kim {\em et~al.\/}}(2022{a}){Kim, Chung, Choi, Succi, Conklin, Longo,
  Ackman, Little, Petranovic, Kalra, Lev, {\rm and} Do}]{RN163}
{Kim D, Chung J, Choi J, Succi MD, Conklin J, Longo MGF, Ackman JB, Little BP,
  Petranovic M, Kalra MK {\em et~al\/}}. 2022{a}. Accurate auto-labeling of
  chest x-ray images based on quantitative similarity to an explainable ai
  model. Nature Communications. 13:1867.

\bibitem[{Kim {\em et~al.\/}}(2022{b}){Kim, Chen, Wang, Mulvey, Yang, Wun,
  Antman-Passig, Luo, Cho, Long-Roche, Ramanathan, Jagota, Zheng, Wang, {\rm
  and} Heller}]{RN154}
{Kim M, Chen C, Wang P, Mulvey JJ, Yang Y, Wun C, Antman-Passig M, Luo HB, Cho
  S, Long-Roche K {\em et~al\/}}. 2022{b}. Detection of ovarian cancer via the
  spectral fingerprinting of quantum-defect-modified carbon nanotubes in serum
  by machine learning. Nature Biomedical Engineering. 6:267--275.

\bibitem[{Kruczkowski {\em et~al.\/}}(2022){Kruczkowski, Drabik-Kruczkowska,
  Marciniak, Tarczewska, Kosowska, {\rm and} Szczerska}]{RN158}
{Kruczkowski M, Drabik-Kruczkowska A, Marciniak A, Tarczewska M, Kosowska M,
  Szczerska M}. 2022. Predictions of cervical cancer identification by photonic
  method combined with machine learning. Scientific Reports. 12:3762.

\bibitem[{Laguarta {\em et~al.\/}}(2020){Laguarta, Hueto, {\rm and}
  Subirana}]{RN143}
{Laguarta J, Hueto F, Subirana B}. 2020. Covid-19 artificial intelligence
  diagnosis using only cough recordings. IEEE Open Journal of Engineering in
  Medicine and Biology. 1:275--281.

\bibitem[{Marks {\em et~al.\/}}(2022){Marks, Jin, Sturman, von Ziegler,
  Kollmorgen, von~der Behrens, Mante, Bohacek, {\rm and} Yanik}]{RN161}
{Marks M, Jin Q, Sturman O, von Ziegler L, Kollmorgen S, von~der Behrens W,
  Mante V, Bohacek J, Yanik MF}. 2022. Deep-learning-based identification,
  tracking, pose estimation and behaviour classification of interacting
  primates and mice in complex environments. Nature Machine Intelligence.
  4:331--340.

\bibitem[{Negi {\em et~al.\/}}(2021){Negi, Yang, Speyer, Pulgarin, Handen,
  Zhao, Tai, Tang, Culley, Yu, Forsythe, Gorelova, Watson, Aaraj, Satoh,
  Sharifi-Sanjani, Rajaratnam, Sembrat, Provencher, Yin, Vargas, Rojas, Bonnet,
  Torrino, Wagner, Schreiber, Dai, Bertero, Ghouleh, Kim, {\rm and}
  Chan}]{RN153}
{Negi V, Yang J, Speyer G, Pulgarin A, Handen A, Zhao J, Tai YY, Tang Y, Culley
  MK, Yu Q {\em et~al\/}}. 2021. Computational repurposing of therapeutic small
  molecules from cancer to pulmonary hypertension. Science Advances.
  7:eabh3794.

\bibitem[{Ouassil {\em et~al.\/}}(2022){Ouassil, Pinals, Bonis-O’Donnell,
  Wang, {\rm and} Landry}]{RN146}
{Ouassil N, Pinals RL, Bonis-O’Donnell JTD, Wang JW, Landry MP}. 2022.
  Supervised learning model predicts protein adsorption to carbon nanotubes.
  Science Advances. 8:eabm0898.

\bibitem[{Sadasivuni {\em et~al.\/}}(2022){Sadasivuni, Saha, Bhatia, Banerjee,
  {\rm and} Sanyal}]{RN160}
{Sadasivuni S, Saha M, Bhatia N, Banerjee I, Sanyal A}. 2022. Fusion of fully
  integrated analog machine learning classifier with electronic medical records
  for real-time prediction of sepsis onset. Scientific Reports. 12:5711.

\bibitem[{Wang {\em et~al.\/}}(2021){Wang, Yao, Gong, Lu, Pang, Li, Yuan, Song,
  Liu, Jin, Ma, Yang, Nie, Zhang, Meng, Zhou, Zhao, Qiu, Zhao, Jiang, Zeng,
  Guo, {\rm and} Yin}]{RN137}
{Wang G, Yao H, Gong Y, Lu Z, Pang R, Li Y, Yuan Y, Song H, Liu J, Jin Y {\em
  et~al\/}}. 2021. Metabolic detection and systems analyses of pancreatic
  ductal adenocarcinoma through machine learning, lipidomics, and multi-omics.
  Science Advances. 7:eabh2724.

\bibitem[{Wood {\em et~al.\/}}(2018){Wood, White, Georgiadis, Emburgh,
  Parpart-Li, Mitchell, Anagnostou, Niknafs, Karchin, Papp, McCord, LoVerso,
  Riley, Diaz, Jones, Sausen, Velculescu, {\rm and} Angiuoli}]{RN149}
{Wood DE, White JR, Georgiadis A, Emburgh BV, Parpart-Li S, Mitchell J,
  Anagnostou V, Niknafs N, Karchin R, Papp E {\em et~al\/}}. 2018. A machine
  learning approach for somatic mutation discovery. Science Translational
  Medicine. 10:eaar7939.

\bibitem[{Xiong {\em et~al.\/}}(2015){Xiong, Alipanahi, Lee, Bretschneider,
  Merico, Yuen, Hua, Gueroussov, Najafabadi, Hughes, Morris, Barash, Krainer,
  Jojic, Scherer, Blencowe, {\rm and} Frey}]{RN148}
{Xiong HY, Alipanahi B, Lee LJ, Bretschneider H, Merico D, Yuen RKC, Hua Y,
  Gueroussov S, Najafabadi HS, Hughes TR {\em et~al\/}}. 2015. The human
  splicing code reveals new insights into the genetic determinants of disease.
  Science. 347:1254806.

\bibitem[{Yaari {\em et~al.\/}}(2021){Yaari, Yang, Apfelbaum, Cupo, Settle,
  Cullen, Cai, Roche, Levine, Fleisher, Ramanathan, Zheng, Jagota, {\rm and}
  Heller}]{RN147}
{Yaari Z, Yang Y, Apfelbaum E, Cupo C, Settle AH, Cullen Q, Cai W, Roche KL,
  Levine DA, Fleisher M {\em et~al\/}}. 2021. A perception-based nanosensor
  platform to detect cancer biomarkers. Science Advances. 7:eabj0852.

\bibitem[{Yala {\em et~al.\/}}(2021){Yala, Mikhael, Strand, Lin, Smith, Wan,
  Lamb, Hughes, Lehman, {\rm and} Barzilay}]{RN151}
{Yala A, Mikhael PG, Strand F, Lin G, Smith K, Wan YL, Lamb L, Hughes K, Lehman
  C, Barzilay R}. 2021. Toward robust mammography-based models for breast
  cancer risk. Science Translational Medicine. 13:eaba4373.

\end{thebibliography}

\end{document}